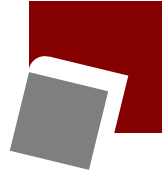

**Institute for Advanced Studies in Basic Sciences**
Gava Zang, Zanjan, Iran

Physics Department

M. Sc. Thesis

Mathematical Physics

Topic
**Matrix Model for membrane and dynamics of D-Particles in a curved space-time geometry and presence of form fields**

By
Qasem Exirifard

Supervisor
Amir Hossein Fatollahi

September 2002

## Abstract

We study dynamics of a membrane and its matrix regularisation. We present the matrix regularisation for a membrane propagating in a curved space-time geometry in the presence of an arbitrary 3-form field. In the matrix regularisation, we then study the dynamics of D-particles. We show how the Riemann curvature of the target space-time geometry, or any other form fields can polarise the D-Particles, cause entanglement among them and create fuzzy solutions. We review the fuzzy sphere and we present fuzzy hyperbolic and ellipsoid solutions.

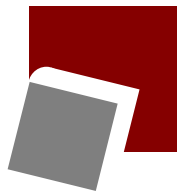

**وزارت علوم، تحقیقات و فناوری**
**دانشگاه تحصیلات تکمیلی علوم پایه**
**گاوازنگ - زنجان**

دانشکدهٔ فیزیک

پایان‌نامه کارشناسی ارشد
گرایش ریاضی-فیزیک

# مدل ماتریسی برای غشا و دینامیک حالتهای مقید $D$-ذره‌ها در فضا-زمان خمیده و حضور میدانهای فُرمی

نگارش
قاسم اکسیری‌فرد

استاد راهنما
دکتر امیر حسین فتح‌اللهی

شهریور ۱۳۸۱

تقدیم به پدر و مادر عزیزم

که در تمام زندگی پشتیبان و یاریگرم

در رسیدن به اهداف مقدسم ادب و دانش هستند

# چکیده

در این پایان‌نامه به بررسی‌ی دینامیک غشاها در پیمانه‌ی مخروط نور و همچنین هنجارش ماتریسی آنها می‌پردازیم. مدل ماتریسی برای غشا دل‌خواه را در فضا-زمان خمیده و حضور میدانهای پیمانه‌ای ارایه می‌دهیم. در هنجارش ماتریسی غشا، به بررسی دینامیک $D$-ذره‌ها می‌پردازیم. نشان می‌دهیم که چگونه تانسور خمش فضا-زمان، یا میدانهای مختلف دیگر می‌توانند قطبشی به $D$-ذره‌ها القا کنند و با این کار بستری بوجود آورند تا $D$-ذره‌ها در هم‌تنیده شوند و حالتهای ناجابه‌جایی و مه‌گون آفریده شود. حل کره‌ی مه‌گون برای $D$-ذره‌ها را مرور می‌کنیم و دو حل جدید بیضی‌گون مه‌گون و هذلولوی‌ی مه‌گون نیز ارایه می‌دهیم.

# فهرست مطالب

# فصل ۱

# کنش غشا در در پیمانه‌ی مخروط نور

در بخش اول این فصل دینامیک غشا بوزونی را در فضا-زمان دلخواه بررسی می‌کنیم. در بخش دوم دینامیک غشا بوزونی را در حضور میدان سه فرمی بررسی خواهیم کرد.

## ۱.۱   کنش غشا در غیاب میدان سه فرمی

به نام گسترشی طبیعی از نظریه‌ی ریسمان و ابژه‌ای که به خودی خود زیباست و در آن هندسه و نسبیت و کوانتم مکانیک به هم مربوط می‌شوند می‌توان تحرک پوسته‌ی بی‌جرم را بررسی کرد. به این پوسته غشا می‌گوییم.

غشا در هنگام حرکت در فضا، یک حجم سه بعدی را در فضا-زمان جاروب می‌کند. این حجم را از این پس جهان‌حجم غشا می‌نامیم. تحرک غشا در فضا-زمان $D+1$ بعدی $x^0, \cdots, x^D$ را با متریک ریمانی $g_{\mu\nu}(x)$ در نظر می‌گیریم. فضا-زمان را از این پس فضای هدف می‌نامیم. جهان‌حجم را با نمادهای $\sigma^0$، $\sigma^1$ و $\sigma^2$ پارامتربندی می‌کنیم. متریک فضای هدف، متریک زیر را بر روی جهان‌حجم القا می‌کند:

$$ds^2 = g_{\mu\nu}dx^\mu dx^\nu = h_{\alpha\beta}d\sigma^\alpha d\sigma^\beta \rightarrow h_{\alpha\beta} = g_{\mu\nu}\frac{\partial x^\mu}{\partial \sigma^\alpha}\frac{\partial x^\nu}{\partial \sigma^\beta}$$

ساده‌ترین کنشی که مستقل از چگونه پارامتربندی کندن جهان حجم باشد و ناوردای لورنتس در فضای هدف باشد عبارت

است از

$$S = -T \int d^3\sigma \sqrt{-\det h} \qquad (1.1)$$

ثابت T انرژی بر واحد سطح غشا است و کنش بالا کنش نامبُ-گُتُ [1] نامیده می‌شود. کنش نامبُ-گُتُ $D+1$ میدان نرده‌ای $x^0, \cdots, x^D$ را با تقارنِ بیرونی $SO(1,D)$ روی فضای سه بعدی جهان حجم با تقارن داخلی موضعی $SO(1,2)$ توصیف می‌کند.

وردش اول کنش نامبُ-گُتُ به متغیرهای $x^D, \cdots, x^0$ معادلات حرکت این کمیتها را می‌دهد:

$$\frac{\delta S}{\delta x_\nu} = 0 \rightarrow \frac{1}{\sqrt{h}}\partial_\alpha(\sqrt{h}h^{\alpha\beta}\partial_\beta x^\nu) + h^{\alpha\beta}\partial_\alpha x^\xi \partial_\beta x^\eta \Gamma^\nu_{\xi\eta} = 0,$$

که در آن $\Gamma^\nu_{\xi\eta}$ هم‌وستار کریستفر در فضای هدف است. اما باید دقت کرد که تمام معادلات بالا از هم مستقل نیستند. برای نشان دادن این امر اجازه دهید معادلات حرکت را به صورت زیر نوشت:

$$\Pi^\mu_\nu = \delta^\mu_\nu - h^{\alpha\beta}\frac{\partial x^\mu}{\partial\sigma^\alpha}\frac{\partial x^\rho}{\partial\sigma^\beta}g_{\rho\nu},$$

$$\Pi^\mu_\nu(x^\nu_{,\alpha\beta} + \Gamma^\nu_{\rho\sigma}x^\rho_{,\alpha}x^\sigma_{,\beta})h^{\alpha\beta} = 0 \qquad (1.2)$$

که در رابطه‌ی بالا $\Pi^\mu_\nu$ عملگر تصویر بر فضای عمود بر جهان‌حجم غشا است. این روش نوشتن نشان می‌دهد که $D-3$ معادله‌ی حرکت مستقل برای حل وجود دارد که دقیقا برابر با تعداد درجه‌های نوسانی‌ی طولی -درجات آزادی فیزیکی- می‌باشد [4].

به علت وجود $\sqrt{\phantom{x}}$ در کنش نامبُ-گُتُ (1.1) کار کردن با این کنش چندان ساده نیست. بهتر است از شر رادیکال خلاص شویم. برای این کار از کنش معادل دیگری که معروف به کنش پلی‌یاکُف [2] استفاده خواهیم کرد. کنش پلی‌یاکف برای

---

[1] Nambu–Goto
[2] Polyakov

میدان‌های فیزیکی $x^\mu$ و میدانِ کمکی $\gamma^{\alpha\beta}$ بر روی جهان‌حجم به صورت زیر تعریف می‌شود:

$$S = -\frac{T}{2} \int d^3\sigma \sqrt{\gamma}(\gamma^{\alpha\beta}\partial_\alpha x^\mu \partial_\beta x^\nu g_{\mu\nu} - 1). \qquad (3.1)$$

در این کنش معادلات حرکت $\gamma^{\alpha\beta}$ معادلاتی قیدی هستند:

$$\frac{\delta S}{\delta \gamma^{\alpha\beta}} = 0 \to \gamma_{\alpha\beta} = h_{\alpha\beta} \qquad (4.1)$$

اگر این کنش را برای لاگ حرکت $\gamma^{\alpha\beta}$ بنویسیم به کنش نامبُ-گُتُ می‌رسیم:

$$\gamma_{\alpha\beta} = h_{\alpha\beta} \to \frac{1}{2}\sqrt{\gamma}(\gamma^{\alpha\beta}\partial_\alpha x^\mu \partial_\beta x^\nu g_{\mu\nu} - 1) = \sqrt{h} \qquad (5.1)$$

از این پس از کنش پلی‌یاکف استفاده می‌کنیم. در مابقی‌ی این فصل فرض می‌کنیم فضای هدف فضای تخت مینکوفسکی باشد.

همان‌گونه که در ابتدای این بخش گفتیم، کنش غشا کنش تقارن‌هایی دارد. اکنون می‌خواهیم از این تقارن‌ها استفاده کنیم و کنش را کمی ساده‌تر کنیم. در صورتی که توپولوژی‌ی جهان حجم به صورت $R \times \Sigma$ باشد که $\Sigma$ خمینه‌ای ریمانی با توپولوژی‌ی ثابت باشد آن‌گاه می‌توان جهان‌حجم را با کمک کمیت دل‌خواه $\nu$ به گونه‌ای پارامتربندی کرد که متریک $\gamma$ به صورت زیر باشد:

$$\gamma = \begin{pmatrix} -\frac{4}{\nu^2}\det \gamma_{ab} & 0 & 0 \\ 0 & \gamma_{11} & \gamma_{12} \\ 0 & \gamma_{12} & \gamma_{22} \end{pmatrix}, a,b \in \{1,2\}. \qquad (6.1)$$

این انتخاب پیمانه‌ی متعامد نامیده می‌شود. کنش در پیمانه‌ی متعامد به صورت زیر ساده می‌شود

$$S = \frac{T\nu}{4} \int d^3\sigma (\dot{x}^\mu \dot{x}_\mu - \frac{4}{\nu^2}\det \gamma_{ab}) \qquad (7.1)$$

که بر حسب براکت پواسون بر روی جهان‌حجم به صورت زیر بیان می‌شود

$$S = \frac{T\nu}{4} \int d^3\sigma (\dot{x}^\mu \dot{x}_\mu - \frac{2}{\nu^2}\{x^\mu, x^\nu\}\{x_\mu, x_\nu\}), \tag{8.1}$$

$$\{x^\mu, x^\nu\} = \epsilon^{ab}\partial_a x^\mu \partial_b x^\nu.$$

معادلات حرکت را نیز می‌توان بر حسب براکت پواسون نمایش داد:

$$\ddot{x}^\mu = \frac{4}{\nu^2}\{\{x^\mu, x^\nu\}, x_\nu\} \tag{9.1}$$

قیدهای سیستم (4.1) به صورت زیر بیان می‌شوند

$$\dot{x}^\mu \partial_a x_\mu = 0, \tag{10.1}$$

$$\dot{x}^\mu \dot{x}_\mu = -\frac{2}{\nu^2}\{x^\mu, x^\nu\}\{x_\mu, x_\nu\}, \tag{11.1}$$

دقت کنید که قیدهای بالا بازپارمتری کردن در راستای $\tau = \sigma^0$ را دست نمی‌زنند. با انتخاب

$$x^\pm = \frac{x^0 \pm x^{D-1}}{\sqrt{2}}, \tag{12.1}$$

$$x^+ = \tau, \tag{13.1}$$

در مختصات مخروط نور (12.1) پیمانه‌ی مخروط نور (13.1) را انتخاب می‌کنیم. در این پیمانه قیدها به صورت صریح حل می‌شوند و داریم

$$\dot{x}^- = \frac{1}{2}\dot{x}^i \dot{x}^i + \frac{1}{\nu^2}\{x^i, x^j\}\{x^i, x^j\} , \ i,j = 1, \cdots D-2 \tag{14.1}$$

دقت کنید که از این جا به بعد تفاوتی بین شاخص‌های بالا و پایین وجود ندارد. اکنون می‌توان هامیلتونی غشا را با محاسبه‌ی

تکانه‌های وابسته به $x^i$ به دست آورد [۴، ۱۱]

$$H = \frac{\nu T}{4} \int d^2\sigma \left(P_i P_i + \frac{2}{\nu^2}\{x^i, x^j\}\{x^i, x^j\}\right), \quad (1.15)$$

$$P_i = \dot{x}_i, \quad$$

و قیدها در فرمول‌بندی‌ی همیلتونی به صورت زیر ترجمه می‌شوند:

$$\{x^i, x^j\} = 0, \quad (1.16)$$

$$\oint x^i \partial x^i = 0, \quad (1.17)$$

قید انتگرالی باید بر روی هر مسیر بسته‌ای درست اعمال شود و قید دیگر معروف به قید گاوس می باشد. اگر بتوان مسیر دلخواه را به طور پیوسته و همواری به یک نقطه فشرده کرد شرطِ انتگرالی بدیهی می‌شود. بنابراین تنها باید مسیرهایی را در نظر گرفت که نتوان به یک نقطه فشرده کرد.

اکنون به تقارنهای همیلتونی با دقت بیشتری می‌خواهیم نگاه کنیم. اگر غشا را با متغیرهای $f^1(\sigma), f^2(\sigma)$ به جای متغیرهای $\sigma^1, \sigma^2$ باز پارامتربندی کنیم چنان که توابع $f$ نگاشتهایی خوش‌رفتار از غشا به غشا باشند آنگاه براکت پواسون به صورت زیر تغییر می‌کند

$$\{x^i, x^j\}_\sigma = \{x^i, x^j\}_f \det \frac{\partial f}{\partial \sigma} \quad (1.18)$$

و هم‌چنین

$$d^2\sigma = d^2 f \det \frac{\partial \sigma}{\partial f} \quad (1.19)$$

۵

بنابراین هامیلتونی به صورت زیر تغییر می‌کند

$$H = \int d\tau \int d^2\sigma (\det \frac{\partial \sigma}{\partial f} \dot{x}^i \dot{x}_i + \det \frac{\partial f}{\partial \sigma} \{x^i, x^j\}_f \{x^i, x^j\}_f) \qquad (20.1)$$

به سادگی می‌توان دید که قید گاوس تغییر نمی‌کند. پس اگر

$$\det \frac{\partial f}{\partial \sigma} = 1 \qquad (21.1)$$

آنگاه هامیلتونی تغییر نمی‌کند. این به معنی است که نگاشت‌های مساحت-نگهدار متصل به یک هامیلتونی‌ی غشا در مخروط نور را تغییر نمی‌دهند.

در حالتی که غشا برآمدگی‌های تندی نداشته باشد در فضای چهار بعدی می‌توان آن را به صورت زیر پارامتربندی کرد:

$$x^0 = t \qquad (22.1\text{الف})$$

$$x^1 = x \qquad (22.1\text{ب})$$

$$x^2 = y \qquad (22.1\text{ج})$$

$$x^3 = h(x, y, t) \qquad (22.1\text{د})$$

این پارامتربندی معروف به پیمانه‌ی مُنژ [3] یا پیمانه‌ی ایستا است. متریک القا شده بر روی جهان‌حجم غشا در این پیمانه عبارت می‌شود از

$$h_{\alpha\beta} = \begin{pmatrix} -1 + \dot{h}^2 & h_{,t}h_{,x} & h_{,t}h_{,y} \\ h_{,t}h_{,x} & 1 + h_{,x}^2 & h_{,x}h_{,y} \\ h_{,t}h_{,y} & h_{,x}h_{,y} & 1 + h_{,y}^2 \end{pmatrix} \qquad (23.1)$$

---

[3] Monge

و

$$-\det h = 1 - \dot{h}^2 + h_{,x}^2 + h_{,y}^2. \tag{1.24}$$

پس کنش غشا در پیمانه‌ی ایستا عبارت می‌گردد از

$$S = -T \int dt\, dx\, dy\, \sqrt{1 + h_{,\alpha} h^{,\alpha}} \tag{1.25}$$

این کنش مشهور به کنشِ بُرن-اینفلد [4] برای میدان نرده‌ای $h(x,y,t)$ می‌باشد. از مدت‌ها پیش این کنش‌ها توسط برن، اینفلد و هایزنبرگ مطالعه شده‌اند [2]. در بخش‌های آتی این دسته کنش‌ها را بیشتر واکاوی می‌کنیم. برای غشا تقریبا ایستا می‌توان از $\dot{h}$ در مقابل $h_{,x}, h_{,y}$ صرف نظر کرد و به دست آورد:

$$S = -T \int dt\, dx\, dy \sqrt{1 + (\nabla h)^2} \;=\; -T \int d\tau\, A(t) \tag{1.26}$$

که در رابطه‌ی بالا $A(t)$ مساحت غشا در زمان $t$ است. همان‌گونه که انتظار داشتیم کنش در این حد خاص نیز چیزی بیش از سطح غشا نیست.

## ۱.۲  کنش غشا در حضور میدان سه-فرمی

در بودِ میدانِ سه-فرمیِ $A_{\mu\nu\rho}$ غشا می‌تواند به صورت بارالکتریکی به این میدان جفت شود:

$$\begin{aligned}
S &= -T \int d^3\sigma \left( \sqrt{-\det h} + \epsilon^{abc} \partial_a x^\mu \partial_b x^\nu \partial_c x^\rho A_{\mu\nu\rho}(x) \right) \\
&= -T \int d^3\sigma \left( \sqrt{-\det h} + 6\, \dot{x}^\mu \partial_1 x^\nu \partial_2 x^\rho A_{\mu\nu\rho}(x) \right)
\end{aligned} \tag{1.27}$$

---

Born–Infeld[4]

۷

معادلات حرکت برای کنش بالا عبارت می‌شود از

$$\frac{1}{\sqrt{h}}\partial_\alpha(\sqrt{h}h^{\alpha\beta}\partial_\beta x^\nu + h^{\alpha\beta}\partial_\alpha x^\xi \partial_\beta x^\eta \Gamma^\nu_{\xi\eta} \; +$$

$$+ (3A^\mu_{\nu\rho,\eta} - A_{\eta\nu\rho,}{}^\mu)\partial_\alpha x^\nu \partial_\beta x^\rho \partial_\gamma x^\eta \epsilon^{\alpha\beta\gamma} \;=\; 0. \tag{۲۸.۱}$$

با معرفی میدان کمکی $\gamma^{\alpha\beta}$ کنش (۲۷.۱) را می‌توان به صورت چندجمله‌ای بر حسب $\partial x$ ها نوشت:

$$S = -\frac{T}{۲}\int d^۳\sigma \sqrt{\gamma}\left((\gamma^{\alpha\beta}\partial_\alpha x^\mu \partial_\beta x^\nu g_{\mu\nu} - ۱) + ۱۲\dot{x}^\mu \partial_۱ x^\nu \partial_۲ x^\rho A_{\mu\nu\rho}(x)\right) \tag{۲۹.۱}$$

که در فضای هدفِ تخت این کنش به صورت زیر ساده می‌شود

$$S = -\frac{T}{۲}\int d^۳\sigma \sqrt{\gamma}\left((\gamma^{\alpha\beta}\partial_\alpha x^\mu \partial_\beta x_\mu - ۱) + ۱۲\dot{x}^\mu \partial_۱ x^\nu \partial_۲ x^\rho A_{\mu\nu\rho}(x)\right) \tag{۳۰.۱}$$

در صورتی که توپولوژی‌ی جهان‌حجم به صورت $R \times \Sigma$ باشد که در $\Sigma$ خمینه‌ای ریمانی با توپولوژی‌ی ثابت است در پیمانه‌ی متعامد (۶.۱) به دست می‌آوریم:

$$S = \frac{T\nu}{۴}\int d^۳\sigma(\dot{x}^\mu \dot{x}_\mu - \frac{۴}{\nu^۲}\det\gamma) - ۶T\int d^۳\sigma \dot{x}^\mu \partial_۱ x^\nu \partial_۲ x^\rho A_{\mu\nu\rho}(x) \tag{۳۱.۱}$$

۸

این کنش در مختصات مخروط نور (۱۲.۱) در پیمانه‌ی مخروط نور (۱۳.۱) عبارت می شود از

$$S = \frac{T\nu}{\mathfrak{r}} \int d^{\mathfrak{r}}\sigma(\dot{x}^i\dot{x}^i - \mathfrak{r}\dot{x}^- - \frac{\mathfrak{r}}{\nu^{\mathfrak{r}}}\det\gamma) - \mathfrak{r}T \int d^{\mathfrak{r}}\sigma \dot{x}^i \partial_{\mathfrak{r}} x^j \partial_{\mathfrak{r}} x^k A_{ijk}(x)$$

$$-\mathfrak{r}T \int d^{\mathfrak{r}}\sigma(\partial_{\mathfrak{r}} x^i \partial_{\mathfrak{r}} x^j A_{+ij}(x) + \dot{x}^- \partial_{\mathfrak{r}} x^i \partial_{\mathfrak{r}} x^j A_{-ij}(x))$$

$$-\mathfrak{r}T \int d^{\mathfrak{r}}\sigma(\partial_{\mathfrak{r}} x^- \partial_{\mathfrak{r}} x^i A_{+-i}(x) + \partial_{\mathfrak{r}} x^i \partial_{\mathfrak{r}} x^- A_{+i-}(x))$$

$$-\mathfrak{r}T \int d^{\mathfrak{r}}\sigma(\dot{x}^i \partial_{\mathfrak{r}} x^- \partial_{\mathfrak{r}} x^j A_{i-j} + \dot{x}^i \partial_{\mathfrak{r}} x^j \partial_{\mathfrak{r}} x^- A_{ij-}(x)) \tag{۳۲.۱}$$

با مرتب کردن جملات بالا می‌رسیم به

$$S = \frac{T\nu}{\mathfrak{r}} \int d^{\mathfrak{r}}\sigma(\dot{x}^i\dot{x}^i - \mathfrak{r}\dot{x}^- - \frac{\mathfrak{r}}{\nu^{\mathfrak{r}}}\det\gamma) - \mathfrak{r}T \int d^{\mathfrak{r}}\sigma \dot{x}^i \partial_{\mathfrak{r}} x^j \partial_{\mathfrak{r}} x^k A_{ijk}(x)$$

$$-\mathfrak{r}T \int d^{\mathfrak{r}}\sigma(\partial_{\mathfrak{r}} x^i \partial_{\mathfrak{r}} x^j A_{+ij}(x) + \dot{x}^- \partial_{\mathfrak{r}} x^i \partial_{\mathfrak{r}} x^j A_{-ij}(x))$$

$$-\mathfrak{r}T \int d^{\mathfrak{r}}\sigma(\{x^-, x^i\} A_{+-i}(x) + \dot{x}^i \{x^-, x^j\} A_{+i-}(x)) \tag{۳۳.۱}$$

معادله‌های قیدی‌ی (۱۰.۱) و (۱۱.۱) عوض نمی‌شوند و با استفاده از آنها داریم:

$$\partial_a x^- = \dot{x}^k \partial_a x_k \rightarrow \{x^-, x^j\} = \dot{x}^k \{x_k, x^j\} \tag{۳۴.۱}$$

با استفاده از رابطه‌ی بالا و (۱۴.۱) می‌توان باز هم کنش را ساده‌تر کرد:

$$S = \frac{T\nu}{\mathfrak{r}} \int d^{\mathfrak{r}}\sigma(\dot{x}^i\dot{x}^i - \frac{\mathfrak{r}}{\nu^{\mathfrak{r}}}\det\gamma)$$

$$-\mathfrak{r}T \int d^{\mathfrak{r}}\sigma(\dot{x}^i \partial_{\mathfrak{r}} x^j \partial_{\mathfrak{r}} x^k A_{ijk}(x) + \partial_{\mathfrak{r}} x^i \partial_{\mathfrak{r}} x^j A_{+ij}(x))$$

$$-\mathfrak{r}T \int d^{\mathfrak{r}}\sigma(\frac{\mathfrak{r}}{\mathfrak{r}}\dot{x}^k \dot{x}^k + \frac{\mathfrak{r}}{\nu^{\mathfrak{r}}}\{x^k, x^l\}\{x^k, x^l\}) \partial_{\mathfrak{r}} x^i \partial_{\mathfrak{r}} x^j A_{-ij}(x)$$

$$-\mathfrak{r}T \int d^{\mathfrak{r}}\sigma(\dot{x}^j \{x^j, x^i\} A_{+-i}(x) + \dot{x}^i \dot{x}^k \{x^k, x^j\} A_{+i-}(x)) \tag{۳۵.۱}$$

۹

پیمانه‌ی زیر را برای میدان سه فرمی انتخاب می‌کنیم

$$A_{-+i} = A_{-ij} = 0 \qquad (36.1)$$

دقت کنید که حق انتخاب چنین پیمانه‌ای را داریم. چون فیزیک تحت تبدیل پیمانه‌ای

$$A_{\mu\nu\lambda} \to A_{\mu\nu\lambda} + \partial_\mu \Lambda_{\nu\rho} + \partial_\nu \Lambda_{\rho\mu} + \partial_\rho \Lambda_{\mu\nu}$$

که در آن $\Lambda$ یک دو-فرم است عوض نمی‌شود. تعداد درجه‌های آزادی این دو-فرم برابر است با $\frac{D(D-1)}{2}$. با انتخاب خوبِ $\Lambda$ می‌توانیم این تعداد از مولفه‌های سه فرم را صفر بگذاریم. با انتخاب $A_{-ij} = 0$ تعداد $\frac{(D-2)(D-3)}{2}$ و با انتخاب $A_{-+i} = 0$ تعداد $D$ متغییر را صفر می‌گذاریم. یعنی در مجموع $D - 3 + \frac{D(D-1)}{2}$ و چون $2 > D$ انتخابمان مجاز هست.

با انتخاب (۳۶.۱) کنش به شکل ساده‌ی زیر تبدیل می‌شود:

$$\begin{aligned}S &= \frac{T\nu}{4} \int d^3\sigma (\dot{x}^i \dot{x}_i - \frac{4}{\nu^2} \det \gamma) \\ &\quad - 3T \int d^3\sigma \dot{x}^j \{x^i, x^k\} A_{ijk}(x) \\ &\quad - 3T \int d^3\sigma \{x^i, x^j\} A_{+ij}(x) \qquad (37.1)\end{aligned}$$

و هامیلتونی عبارت می‌شود از

$$\begin{aligned}H &= \frac{T\nu}{4} \int d^2\sigma (\dot{x}^i \dot{x}^i + \frac{4}{\nu^2} \{x^i, x^j\}\{x^i, x^j\}) + 3T \int d^2\sigma \{x^i, x^j\} A_{+ij} \\ P_i &= \dot{x}_i - 3\{x^j, x^k\} A_{ijk}(x) \qquad (38.1)\end{aligned}$$

که در نبود میدان سه‌فرمی تبدیل به (۱۵.۱) می‌شود.

## 3.1 ناپایداری غشا و رشد مو بر آن

انرژی یک غشا تقریبا ایستا برابر است با سطح آن. این انرژی باعث می‌شود که دو غشا که تنها تفاوتشان در وجود تعدادی استوانه‌ی بسیار نازک مو شکل می‌باشند در سطح مکانیک کلاسیکی انرژی یکسانی داشته باشند و در نتیجه تبهگنی انرژی بسیار زیاد است.

در سطح مکانیک کوانتمی به سبب اصل نبود قطعیت هایزنبرگ مساله فرق می‌کند و دو غشا دیگر انرژی یکسانی ندارند. برای بهتر دیدن این موضوع دو غشا تخت که بر روی یکی استوانه‌ای بسیار نازک به شعاع قاعده‌ی $a$ و ارتفاع $H$ نصب شده است را به صورت نیمه کلاسیکی در نظر می‌گیریم. تفاوت مساحت این دوغشا عبارت می‌شود از

$$\Delta S = \pi a H$$

در حد $a \to 0$ این اختلاف قابل صرف نظر کردن است. حال فرض کنید تحرک استوانه همانند جسمی صلب باشد و کل استوانه را یک ذره در نظر بگیرید. با تعیین جای استوانه مکان این شبه ذره را با دقت $2a$ تعیین کرده‌ایم. در نتیجه به خاطر اصل نبود قطعیت هایزنبرگ عدم قطعیتی برابر با $\Delta P = \frac{\hbar}{4a}$ در تکانه‌ی استوانه در دو راستای موازی با سطح غشا داریم. پس کمینه‌ی انرژی جنبشی استوانه عبارت است از

$$E_k = (\frac{\hbar}{4a})^2 M, \ M = 2\pi a H$$

که برابر است با

$$E_k = \frac{\pi \hbar^2 H}{8a}$$

پس بود استوانه‌ی نازک در کل به اندازه‌ی

$$E_k = 2\pi a H + \frac{\pi \hbar^2 H}{8a}$$

انرژی‌ی سیستم را افزایش می‌دهد. اگر شعاع استوانه را به سمت صفر میل دهیم تفاوت انرژی به صفر نمی‌رسد. پس مکانیک کوانتمی تبهگنی در انرژی را می‌کاهد. در پایان فصل آینده این موضوع را در هنجارش ماتریسی‌ی غشا مطالعه می‌کنیم و به علاوه نشان می‌دهیم که تبهگنی‌ی انرژی ابر غشا بر عکس غشا در سطح کلاسیکی و کوانتمی یکسان است.

# فصل ۲

# مدل ماتریسی برای غشا

آقای هُپ [1] و گلدستن [2] در سال ۱۹۸۲ یک روش هوش‌مندانه برای هنجارش غشای کروی یافتند [۸] . آنها نشان دادند که پایه‌هایی برای نمایش یکانی گروه $SU(N)$ در حد $N$ های بزرگ می‌توان انتخاب کرد که جبر لی‌ی آن همسان با جبر لی‌ی گروه دیفیومرفیزم‌های مساحت-نگه‌دار متصل به یک برای یک کره باشد. با استفاده از این مطلب / آنها مدلی ماتریسی برای غشا کروی نوشتند.

در این فصل ما این فرایند را برای غشا چنبره‌ای شکل در چاربعد انجام می‌دهیم. ره‌یافت را یه به سه بخش خُرد می‌کنیم:

۱. در ابتدا هامیلتونی‌ی غشا را بر حسب تبدیلات فوریه می‌نویسیم.

۲. پایه‌ای مناسب بر جبر لی‌ی گروه $SU(N)$ در نمایش هم‌یوغ آن انتخاب می‌کنیم.

۳. با استفاده از قسمت‌های یک و دو, یک هامیلتونی با تقارن بیرونی‌ی $SU(N)$ که متناظر با هامیلتونی‌ی غشا در مخروط نور است را می‌نویسیم.

---

Hope Jens [1]
Goldston Jeffery [2]

۱۳

## ۲.۱  تبدیلات فوریه‌ی هامیلتونی غشا چنبره‌ای در چاربعد

یک غشا چنبره‌ای شکل ثابت را در چاربعد به کمک نمادهای $\theta$ و $\phi$ می‌توان به صورت زیر پارامتربندی کرد:

$$
\begin{aligned}
x &= (a + b\sin\theta)\cos\phi \\
y &= (a + b\sin\theta)\sin\phi \\
z &= b\cos\theta
\end{aligned}
\qquad (\text{۲.۱})
$$

در حالتی که غشا نوسانهای کمی حول این حالت سکون داشته باشد باز هم می‌توان تمام ِغشا را با پارامترهای $\theta$ و $\phi$ پوشاند. البته در این حالت خواهیم داشت

$$
\begin{aligned}
x &= x(\theta,\phi,t) \\
y &= y(\theta,\phi,t) \\
z &= z(\theta,\phi,t)
\end{aligned}
\qquad (\text{۲.۲})
$$

پس از انتخاب $T = \frac{\nu}{2} = 1$ و $P_x = \dot{x}$ و $P_y = \dot{y}$ هامیلتونی غشا در پیمانه‌ی مخروط نور عبارت می شود از

$$
H[x,y,\dot{x},\dot{y}] = \frac{1}{2}\int_T d\Omega(\dot{x}^2 + \dot{y}^2 + \{x,y\}^2) \qquad (\text{۲.۳الف})
$$

$$
d\Omega = d\theta d\phi \qquad (\text{۲.۳ب})
$$

$$
\{x,y\} = \frac{\partial x}{\partial \theta}\frac{\partial y}{\partial \phi} - \frac{\partial x}{\partial \phi}\frac{\partial y}{\partial \theta} \qquad (\text{۲.۳ج})
$$

و قیدها به صورت زیر خوانده می‌شوند:

$$\{x, \dot{x}\} + \{y, \dot{y}\} = 0 \qquad (4.2)$$

$$\int d\theta (\dot{x}\partial_\theta x + \dot{y}\partial_\theta y) = 0 \qquad (5.2)$$

$$\int d\phi (\dot{x}\partial_\phi x + \dot{y}\partial_\phi y) = 0 \qquad (6.2)$$

اکنون اجازه دهید هامیلتونی را بر حسب تبدیلات فوریه‌ی $x$ و $y$ بنویسیم. یعنی قرار دهیم:

$$x = \sum_{m,n} x_{mn} e^{i(m\phi + n\theta)}, \ x^*_{mn}(t) = x_{(-m)(-n)}(t) \qquad (7.2)$$

$$y = \sum_{m,n} y_{mn} e^{i(m\phi + n\theta)}, \ y^*_{mn}(t) = y_{(-m)(-n)}(t) \qquad (8.2)$$

که با استفاده از عبارتهای بالا سریعا به دست می‌آوریم که

$$\int d\theta d\phi \, \dot{x}^2(t) = 4\pi^2 \sum_{m,n} |\dot{x}_{mn}|^2 \qquad (9.2)$$

$$\int d\theta d\phi \, \dot{y}^2(t) = 4\pi^2 \sum_{m,n} |\dot{y}_{mn}|^2 \qquad (10.2)$$

$$\int d\theta d\phi \, \{x, y\}^2 = -4\pi^2 \sum_{\vec{m},\vec{n},\vec{m}',\vec{n}'} m \times n \, m' \times n' \, x_{\vec{m}} x_{\vec{m}'} y_{\vec{n}} x_{\vec{n}'} \delta_{\vec{m}+\vec{m}'+\vec{n}+\vec{n}',0} \qquad (11.2)$$

و هامیلتونی تبدیل می‌شود به

$$H = 2\pi^2 \sum (|\dot{x}_{mn}|^2 + |\dot{y}_{mn}|^2)$$

$$- 2\pi^2 \sum m \times n \, m' \times n' \, x_{\vec{m}} x_{\vec{m}'} y_{\vec{n}} x_{\vec{n}'} \delta_{\vec{m}+\vec{m}'+\vec{n}+\vec{n}',0} \qquad (12.2)$$

دقت کنید که $\epsilon^{ab}m_a n_b = m \times n$. به این هامیلتونی می‌توان به صورت سیستمی متشکل از بی‌نهایت ذره‌ی اسکالر $x_{mn}(t)$ و $y_{mn}(t)$ با پتانسیل برهم‌کنشی

$$V = -2\pi^2 \sum m \times n m' \times n' x_{\vec{m}} x_{\vec{m}'} y_{\vec{n}} x_{\vec{n}'} \delta_{\vec{m}+\vec{m}'+\vec{n}+\vec{n}',0} \tag{2.13}$$

که تقارن‌های پیچیده‌ای دارد نگاه کرد. کار کردن با بی‌نهایت ذره ساده نیست. اجازه دهید ابتدا خود را به مدهای نوسانی کوچکتر از $N$ محدود کنیم و سپس $N$ را به سمت بی‌نهایت میل دهیم.

ماتریس‌های $N \times N$ ‌ی $T_{mn}$ که در رد‌گیری عمود بر هم هستند و در رابطه‌های زیر صدق می‌کنند را در نظر بگیرید و انتخاب کنید:

$$tr(T_{\vec{m}} T_{\vec{n}}) = \delta_{m+n,0}, \tag{2.14}$$

$$T_{\vec{m}}^\dagger = T_{-\vec{m}}, \tag{2.15}$$

$$[T_{\vec{m}}, T_{\vec{n}}] = im \times n T_{\vec{m}+\vec{n}} \tag{2.16}$$

ماتریس‌های هرمیتی‌ی $X$ و $Y$ را به کمک $T$ ها به صورت زیر تعریف کنید:

$$X = \sum_{m,n} T_{mn} x_{mn} \tag{2.17}$$

$$Y = \sum_{m,n} T_{mn} y_{mn} \tag{2.18}$$

آنگاه ماتریس‌های $X$ و $Y$ در روابط زیر صدق می‌کنند

$$tr(\dot{X}^2) = \sum_{m,n} |\dot{x}_{mn}(t)|^2, \tag{۱۹.۲}$$

$$tr(\dot{Y}^2) = \sum_{m,n} |\dot{y}_{mn}(t)|^2, \tag{۲۰.۲}$$

$$[X,Y] = i\vec{m} \times \vec{n} x_{\vec{m}} y_{\vec{n}} T_{\vec{m}+\vec{n}}, \tag{۲۱.۲}$$

$$[X,Y]^2 = -\vec{m} \times \vec{n} \vec{m}' \times \vec{n}' x_{\vec{m}} y_{\vec{n}} x_{\vec{m}'} y_{\vec{n}'} T_{\vec{m}+\vec{n}} T_{\vec{m}'+\vec{n}'}, \tag{۲۲.۲}$$

با استفاده از (۱۴.۲) و (۲۲.۲) می‌توان پتانسیل (۱۳.۲) را به صورت زیر نوشت:

$$V = \frac{1}{2} tr[X,Y]^2 \tag{۲۳.۲}$$

اکنون هامیلتونی را می‌توان بر حسب ماتریس‌های $X$ و $Y$ نوشت:

$$H = 2\pi^2 tr(\dot{X}^2 + \dot{Y}^2) + 2\pi^2 tr[X,Y]^2 \tag{۲۴.۲}$$

و قید گاوس هم تبدیل می‌شود به

$$[\dot{X}, X] + [\dot{Y}, Y] = 0 \tag{۲۵.۲}$$

بررسی دو شرط انتگرالی دیگر را به بعد موکول می‌کنیم. در روابط بالا ماتریس‌های $T$ با رابطه‌ی جابه‌جایی (۱۶.۲) مولدهای گروه تقارنی‌ی هامیلتونی‌ی (۲۴.۲) می‌باشند. می‌توان با به دست آوردن جبرِ لی‌ی گروه دیفیومرفیزم‌های مساحت‌نگه‌دار متصل به یک به صورت زیر این مطلب را دید.

ناوردا بودن هامیلتونی‌ی (۳.۲) و قیدها تحت دیفیومرفیزم‌های مساحت‌نگه‌دار متصل به یک به این معنی است که بستگی‌ی هامیلتونی (۳.۲) به $(\vec{P}, \vec{x})$ تحت نگاشت هموار $(\theta, \phi)$ به $(\theta', \phi')$ به شرط یک‌بودن ژاکوبین تبدیل تغییر

۱۷

نمی‌کند. برای تحقیقِ درستیِ این مطلب ریز-نگاشت $\theta \to \theta + \delta\theta$ و $\phi \to \phi + \delta\phi$ با شرطِ

$$\left| \begin{array}{cc} \frac{\partial \theta'}{\partial \theta} & \frac{\partial \theta'}{\partial \phi} \\ \frac{\partial \phi'}{\partial \theta} & \frac{\partial \phi'}{\partial \phi} \end{array} \right| = 1$$

را در نظر بگیرید. برای این تبدیلات $\delta\theta = \partial_\phi f$ و $\delta\phi = -\partial_\theta f$. برای تغییرات هر تابع دلخواه $Z(\theta, \phi)$ داریم

$$\delta_f Z = (\partial_\theta f \partial_\phi - \partial_\phi f \partial_\theta) Z \tag{۲۶.۲}$$

پس اثر دیفیومرفیزم مساحت نگهدار $f$ بر روی فضای توابع هموار تعریف شده بر روی سطح غشا را می‌توان با عمل‌گر زیر نمایش داد:

$$\hat{X}_f = \partial_\theta f \partial_\phi - \partial_\phi \partial_\theta \tag{۲۷.۲}$$

دو عمل‌گر دیفیومرفیزم مساحت نگهدار $f$ و $g$ در رابطه‌ی جابه‌جایی زیر صدق می‌کنند:

$$[\hat{X}_f, \hat{X}_g] = \hat{X}_{\{f,g\}} \tag{۲۸.۲}$$

$$\{f, g\} = \partial_\theta f \partial_\phi g - \partial_\theta g \partial_\phi f$$

این مجموعه عمل‌گرها بسته هستند، در رابطه‌ی ژاکوبی صدق می‌کنند، پس یک جبر لی می‌سازند. نمایش این عمل‌گرها در پایه‌ی گسسته‌ی فوریه که پایه‌ای کامل برای توابع تعریف شده روی چنبره است به صورت زیر می‌باشد

$$(\hat{X}_m)_{kl} = \int d^2\sigma e^{-ik.\sigma} \hat{X}_{e^{im.\sigma}} e^{il\sigma} = im \times l \delta_{m, k-l}$$

$$m \times l = \epsilon^{ab} m_a l_b, \tag{۲۹.۲}$$

۱۸

و رابطه‌ی جابه‌جایی (۲۸.۲) در پایه‌ی فوریه به صورت زیر در می‌آید:

$$([\hat{X}_m, \hat{X}_n])_{kl} = im \times n(\hat{X}_{m+n})_{kl} \qquad (۳۰.۲)$$

این رابطه رابطه‌ی جابه‌جایی (۱۶.۲) است [۶].

در بخش بعد نشان می‌دهیم که مولدهایی برای نمایش هم‌یوغ گروه $SU(N)$ می‌توان یافت که در حد $N \to \infty$ رابطه‌های (۱۴.۲) و (۱۵.۲) و (۱۶.۲) برآورده شوند.

## ۲.۲  نمایشهای هم‌یوغ جبر لی‌ی گروه $SU(N)$

در این بخش ابتدا نمایش‌های گروه‌های $SU(N)$ را برای $N$ های فرد و سپس زوج مرور می‌کنیم

### ۲.۲.۱  نمایشهای هم‌یوغ جبر لی‌ی گروه $SU(N)$ برای $N$ های فرد

مولدهای گروه $SU(N)$ برای $N$ های فرد را می‌توان به صورت زیر انتخاب کرد : [۵]

$$Y^{SU(N)}_{(m_۱,m_۲)} = e^{\frac{۲\pi i}{N} m_۱ m_۲} U^{m_۱} V^{m_۲} \qquad (۳۱.۲)$$

که $U$ و $V$ به ترتیب ماتریس‌های زیر می‌باشند:

$$U = \begin{pmatrix} ۱ & ۰ & \cdots & & ۰ \\ ۰ & e^{\frac{۴\pi i}{N}} & \cdots & & ۰ \\ \vdots & \vdots & \ddots & & \vdots \\ ۰ & ۰ & \cdots & & e^{\frac{۴\pi i}{N}(N-۱)} \end{pmatrix}, V = \begin{pmatrix} ۰ & ۱ & ۰ & \cdots & ۰ \\ ۰ & ۰ & ۱ & \cdots & ۰ \\ \vdots & \vdots & \vdots & \ddots & \vdots \\ ۰ & ۰ & ۰ & \cdots & ۱ \\ ۱ & ۰ & ۰ & \cdots & ۰ \end{pmatrix}, \qquad (۳۲.۲)$$

۱۹

$U$ و $V$ هر دو ماتریسهایی از مرتبه‌ی $N$ می‌باشند ($U^N = V^N = 1$) و در رابطه‌ی $UV = e^{\frac{4\pi i}{N}} VU$ صدق می‌کنند.

می‌توان دید

$$Y^{SU(N)}_{m_1+k_1N, m_2+k_2N} = Y^{SU(N)}_{m_1, m_2}, \quad k_i \in Z \tag{33.2}$$

$$Y^{SU(N)}_{\vec{m}} Y^{SU(N)}_{\vec{m}'} = e^{\frac{2\pi i}{N} \vec{m} \times \vec{m}'} Y^{SU(N)}_{\vec{m}+\vec{m}'} \tag{34.2}$$

$$Y^{SU(N)\dagger}_{\vec{m}} = Y^{SU(N)}_{-\vec{m}} \tag{35.2}$$

و رابطه‌ی جابه‌جایی این مولدها به صورت زیر هست

$$[Y^{SU(N)}_m, Y^{SU(N)}_n] = -2i \sin(\frac{2\pi}{N} m \times n) Y^{SU(N)}_{m+n} \tag{36.2}$$

با استفاده از رابطه‌ی بالا مولدهای گروه $SU(N)$ در نمایش هم‌یوغش را می‌توان به صورت زیر انتخاب کرد

$$(X^N_m)_{kl} = -2i \sin(\frac{2\pi}{N} m \times l) \delta_{m, k-l} \tag{37.2}$$

هنگامی که مدهای نوسانی کوچک غشا می‌کنیم می‌توانیم تقریب زیر را به کار ببریم

$$(X^N_m)_{kl} \approx -\frac{4\pi i}{N} m \times l \delta_{m, k-l} \tag{38.2}$$

$$([X^N_m, X^N_n])_{kl} \approx -\frac{4\pi i}{N} m \times n (X^N_{m+n})_{kl}$$

## 2.2.2  نمایشهای هم‌یوغ جبر لی‌ی گروه $SU(N)$ برای $N$ های زوج

مولدهای گروه $SU(N)$ برای $N$ های زوج را می‌توان به صورت زیر انتخاب کرد [5]

$$Y^{SU(N)}_{(m_1, m_2)} = e^{\frac{2\pi i}{N} m_1 m_2} U^{m_1} V^{m_2}$$

۲۰

که $U$ و $V$ به ترتیب ماتریسهای زیر می‌باشند:

$$U = e^{\frac{\pi i}{N}} \begin{pmatrix} 1 & 0 & \cdots & 0 \\ 0 & e^{\frac{2\pi i}{N}} & \cdots & 0 \\ \vdots & \vdots & \ddots & \vdots \\ 0 & 0 & \cdots & e^{\frac{2\pi i}{N}(N-1)} \end{pmatrix}, V = \begin{pmatrix} 0 & 1 & 0 & \cdots & 0 \\ 0 & 0 & 1 & \cdots & 0 \\ \vdots & \vdots & \vdots & \ddots & \vdots \\ 0 & 0 & 0 & \cdots & 1 \\ -1 & 0 & 0 & \cdots & 0 \end{pmatrix}, \quad (2.39)$$

$U$ و $V$ این بار در رابطه‌ی زیر صدق می‌کنند:

$$U^N = V^N = -1 \quad (2.40)$$

و نمایشهای هم‌یوغ را بر حسب پایه‌های زیر می‌توان بیان کرد

$$Y^{SU(N)}_{(m_1,m_2)} = e^{\frac{\pi i}{N}m_1 m_2} U^{m_1} V^{m_2} \quad (2.41)$$

رابطه‌ی جابه‌جایی این پایه‌ها عبارتند از

$$[Y^{SU(N)}_m, Y^{SU(N)}_n] = 2i \sin(\frac{\pi}{N} m \times n) Y^{SU(N)}_{m+n} \quad (2.42)$$

پس برای نمایش هم‌یوغ $SU(N)$ می‌توان پایه‌های ماتریسی زیر را انتخاب کرد

$$\begin{aligned} (X^N_m)_{kl} &= -2i \sin(\frac{\pi}{N} m \times l) \delta_{m,k-l} & (2.43) \\ ([X^N_m, X^N_n])_{kl} &= -2i \sin(\frac{\pi}{N} m \times n)(X^N_{m+n})_{kl} & (2.44) \end{aligned}$$

دوباره در حدِ $N$ های بزرگ برای $m$ و $n$ های کوچک به دست می‌آوریم که

$$(X_m^N)_{kl} = -\frac{2\pi i}{N} m \times l \delta_{m,k-l} \qquad (2.45)$$

$$([X_m^N, X_n^N])_{kl} = -\frac{2\pi i}{N} m \times n (X_{m+n}^N)_{kl} \qquad (2.46)$$

### 2.2.3  دیفیومرفیزمهای مساحت‌نگه‌دار متصل به یک و نمایشهای گروه $SU(N)$

برای $N$ های فرد با مقایسه‌ی (2.30) با (2.38) می‌بینیم که با تبدیل

$$\frac{4\pi}{N}(X_m^N)_{kl} \to (\hat{X}_m)_{kl} \qquad (2.47)$$

رابطه‌ی جابه‌جایی (2.38) به (2.30) یا معادلش (2.16) تبدیل می‌شود. برای $N$ های زوج می‌توان دید که با تبدیل

$$\frac{2\pi}{N}(X_m^N)_{kl} \to (\hat{X}_m)_{kl} \qquad (2.48)$$

رابطه‌ی جابه‌جایی (2.45) به (2.30) تبدیل می‌شود. به طور مستقیم می‌توان نشان داد که در هر دو حالت زوج و فرد پس از تبدیلات (2.47) و (2.48) رابطه‌ی (2.14) برآورده می‌شد. رابطه‌ی (2.15) نیز از (2.35) آشکار است. پس همان‌گونه که در پایان بخش قبل گفتیم گروه تبدیلات دیفیومرفیزمهای مساحت نگه‌دار متصل به یک برای یک چنبره مانند حد $N \to \infty$ نمایش هم‌پیوغ گروه $SU(N)$ است.

## 2.3  مدل ماتریسی برای غشا چنبره‌ای در پیمانه‌ی مخروط نور

در قسمت اول این فصل با کمک ماتریسهایی هامیلتونی غشا را برای مدهای نوسانی پایین به صورت زیر نوشتیم

$$H = 2\pi^2 tr(\dot{X}^2 + \dot{y}^2) + 2\pi^2 tr[X,Y]^2 \qquad (2.49)$$

$$[X, \dot{X}] + [Y, \dot{Y}] = 0 \qquad (2.50)$$

که ماتریس‌های هرمیتی $X$ و $Y$ بوسیله‌ی تبدیل‌های فوریه‌ی $y_{mn}$ و $x_{mn}$ در (۱۷.۲) و (۱۸.۲) تعریف شده‌اند. تنها خصوصیت‌های ماتریس‌های $T$ همان‌طور که در (۱۴.۲) و (۱۶.۲) فرض کردیم عبارتند از

$$tr(T_m T_n) = \delta_{m+n,\circ} \ , [T_m, T_n] = im \times n T_{m+n} \ , \ T_m^\dagger = T_{-m}$$

سپس نشان دادیم که مولدهای گروه $SU(N)$ در نمایش هم‌یوغش را به گونه‌ای می‌توان انتخاب کرد که رابطه‌های بالا برای مدهای نوسانی پایین برآورده شود. بنابراین $T$ها را می‌توان مولدهای گروه‌های $SU(N)$ در نمایش هم‌یوغش در نظر گرفت. در این صورت هامیلتونی گروه تقارن بیرونی $SU(N)$ را دارا می‌باشد. با بازتعریف

$$X \to ۲\pi X \ , \ Y \to ۲\pi Y \tag{۵۱.۲}$$

هامیلتونی تبدیل می‌شود به

$$H = \frac{۱}{۲} tr(\dot{X}^۲ + \dot{y}^۲) + \frac{۱}{۸\pi^۲} tr[X,Y]^۲ \tag{۵۲.۲}$$

$$[X, \dot{X}] + [Y, \dot{Y}] = \circ \tag{۵۳.۲}$$

و در این جا می‌توان دید دو قید انتگرالی (۵.۲) و (۶.۲) خود به خود برآورده می‌شوند. هامیلتونی بالا مشهور به مدل ماتریسی برای غشا می‌باشد. در به دست آوردن هامیلتونی تنها نوسانهایی با طول موج بسیار بیشتر از $\frac{۲\pi}{N}$ را در نظرگرفتیم. به عبارتی ملموس‌تر سطح چنبره را با یک شبکه‌بندی چنبره‌ای جایگزین کرده‌ایم. آقای Florates در [۷] مستقیما چنبره را شبکه‌بندی کرده است و نشان داده است که گروه دیفیومرفیزمهای مساحت‌نگه‌دار متصل به یک برای چنبره‌ای شبکه‌ای شده تبدیل به گروه $SU(N)$ می‌شود. ما کارهای ایشان را مرور و ساده‌سازی کردیم.

۲۳

## ۴.۲ مدل ماتریسی برای چنبره در حضور میدان سه‌فرمی

این بخش نو و تازه است و توسط نگارنده انجام شده است. این اولین مدل ماتریسی برای چنبره در حضور میدان سه‌فرمی است. با توجه به رابطه‌ی (۳۸.۱) هامیلتونی‌ی غشای چنبره‌ای در پیمانه‌ی مخروط نور در فضای هدف چاربعدی با پارامتربندی‌ی (۲.۲) در حضور میدان سه‌فرمی $A_{\mu\nu\eta}$ با تثبیت پیمانه‌ی (۳۶.۱) و انتخاب مرسوم $1 = \frac{\nu}{۲} = t$ عبارت است از

$$H = \frac{1}{۲}\int d^۲\sigma(\dot{x}^۲ + \dot{y}^۲ + \{x,y\}^۲) + ۶\int d^۲\sigma\{x,y\}A_{+xy} \tag{۵۴.۲}$$

$$P_x = \dot{x} - ۶\{x,y\}A_{xxy} \tag{۵۵.۲}$$

$$P_y = \dot{y} - ۶\{x,y\}A_{yxy} \tag{۵۶.۲}$$

در این روابط میدان سه‌فرمی تابعیت دلخواه از فضا-زمان چار بعدی دارد.

اکنون می‌خواهیم با کمک فرایندی مشابه با آن چه که در بخش‌های پیش انجام دادیم هامیلتونی‌ی ماتریس همزادی برای غشا در حضور میدان سه‌فرمی بنویسیم. برای بهتر فهمیدن فرایند اجازه دهید ابتدا مساله را بسیار ساده کنیم و فرض کنیم تنها مولفه‌ی غیر صفر میدان سه‌فرمی به صورت زیر است

$$A_{+xy} = A(t)x$$

در این حالت با استفاده از (۷.۲) و (۸.۲) جمله‌ی برهم‌کنشی با میدان سه‌فرمی در (۵۴.۲) عبارت می‌شود از:

$$-۶\int d^\sigma\{x,y\}A_{+xy} = -۶i \times ۴\pi^۲ A(t)\sum m \times n x_m y_n x_{m'}\delta_{m+m'+n,۰} \tag{۵۷.۲}$$

با کمک ماتریس‌های $T$ و تعریف ماتریس‌های $X$ و $Y$ با (۷.۲) و (۸.۲) می‌بینیم که

$$tr([X,Y]X) = i\sum m \times n x_m y_n x_{m'}\delta_{m+n+n',۰} \tag{۵۸.۲}$$

پس در این حالت خاص تاثیر $A_{+xy}$ بر مدل ماتریسی ارایه شده در (۲۴.۲) اضافه کردن یک جمله به آن به صورت زیر است

$$H = 2\pi^2 tr(\dot{X}^2 + \dot{Y}^2) + 2\pi^2 tr[X,Y]^2 + 4\pi^2 tr\{-6[X,Y]A(t)X\} \qquad (59.2)$$

رابطه‌ی بالا را به صورت زیر هم می‌توان نوشت

$$H = 2\pi^2 tr(\dot{X}^2 + \dot{Y}^2) + 2\pi^2 tr[X,Y]^2 + 4\pi^2 tr\{-6[X,Y]A_{+xy}[X]\} \qquad (60.2)$$

که در آن $A_{+xy}[X]$ تابعی از ماتریس $X$ است که به صورت $A_{+xy}[X] = A(t)X$ تعریف شده است. به عبارتی ساده مختصه‌ی مکانی تبدیل به یک ماتریس شده است.

اگر $A_{+xy}$ را ترکیبی خطی از $x, y, z$ در نظر بگیریم باز هم به همین نتیجه‌ی بالا خواهیم رسید، البته باید ماتریس $Z$ را هم همچون ماتریس‌های $X$ و $Y$ به صورت $Z = \sum z_{mn} T_{mn}$ که $Z_{mn}$ مولفه‌های بسط فوریه‌ی $z(\sigma)$ می‌باشند تعریف نمود.

اکنون یک قدم پیش‌تر می‌رویم و تنها مولفه‌ی غیر صفر پتانسیل سه‌فرم را به صورت زیر در نظر می‌گیریم:

$$A_{+xy} = xyA(t)$$

در این حالت جمله‌ی برهم‌کنشی با میدان سه‌فرمی در (۵۴.۲) با کمک تبدیل فوریه‌ی $x$ و $y$ به صورت زیر می‌تواند نوشته شود:

$$-6\int d^2\sigma\{x,y\}A_{+xy} = -24i\pi^2 A(t)\sum m \times n\, x_m y_n x_{m'} y_{n'} \delta_{m+m'+n+n',0} \qquad (61.2)$$

برای یافتن معادل ماتریسی عبارت بالا ابتدا ماتریسهای زیر را حساب می‌کنیم

$$tr([X,Y]XY) = m \times n x_m y_n x_{m'} y_{n'} e^{\frac{2\pi i}{N} m' \times n'} \delta_{m+m'+n+n',0} \qquad (2.62)$$

$$tr([X,Y]YX) = m \times n x_m y_n x_{m'} y_{n'} e^{-\frac{2\pi i}{N} m' \times n'} \delta_{m+m'+n+n',0} \qquad (2.63)$$

در مدهای نوسانی پایین داریم

$$tr([X,Y]XY) \approx m \times n x_m y_n x_{m'} y_{n'} \delta_{m+m'+n+n',0} (1 - \frac{2\pi i}{N} m' \times n' + O(\frac{1}{N^2})) \qquad (2.64)$$

$$tr([X,Y]YX) \approx m \times n x_m y_n x_{m'} y_{n'} \delta_{m+m'+n+n',0} (1 + \frac{2\pi i}{N} m' \times n' + O(\frac{1}{N^2})) \qquad (2.65)$$

در حد $N$های بزرگ هر دو عبارت بالا آن چه که دوست داریم هستند. اما میانگین آنها بهتر است چون در مدهای نوسانی پایین داریم

$$\frac{1}{2}(tr([X,Y]XY) + tr([X,Y]YX)) \approx m \times n x_m y_n x_{m'} y_{n'} \delta_{m+m'+n+n',0}(1 + O(\frac{1}{N^2}))$$

$$(2.66)$$

و در $N$های کوچکتر می‌توان نتیجه‌ی بهتری به دست آورد. با متوسط گیری ما اثرات طول محدود شبیه‌سازی یا تقریب خود را ضعیف‌تر می‌کنیم و جواب بهتری به دست می‌آوریم. این نکته به این زبان برای اولین بار در این پایان‌نامه دارد گزارش می‌شود. متوسط بر روی ردهای تمام جایگشتهای ممکن را رد متقارن می‌نامیم و با نماد $Str$ نمایش می‌دهیم.

$$Srt[T_1 \cdots T_n] = \frac{\sum_{All\ Permutations} T_{p_1}..T_{p_n}}{Number\ of\ Permutations} \qquad (2.67)$$

با استفاده از (2.66) و (2.61) و (2.24) و رد متقارن، مدل ماتریسی در این حالت به صورت زیر در می‌اید

$$H = 2\pi^2 tr(\dot{X}^2 + \dot{Y}^2) + 2\pi^2 tr[X,Y]^2 + 4\pi^2 Str\{-6[X,Y]A(t)XY\}$$

که با تعریف $A_{+xy}[\vec{X}] = A(t)XY$ می‌رسیم به

$$H = 2\pi^2 tr(\dot{X}^2 + \dot{Y}^2) + 2\pi^2 tr[X,Y]^2 + 4\pi^2 Str\{-6[X,Y]A_{+xy}[\vec{X}]\} \qquad (68.2)$$

در حالتی که $A_{+xy}$ دلخواه باشد و یک بسط تیلور در نزدیکی مبدا به صورت زیر داشته باشد

$$A_{+xy} = \sum A_{\alpha\beta\gamma} x^\alpha y^\beta z^\gamma$$

جمله‌ی برهم‌کنشی با میدان فرمی در (54.2) به صورت زیر می‌تواند نوشته شود

$$-6\int d^2\sigma\{x,y\}A_{+xy} = -24i \sum_{\alpha\beta\gamma} A^{\alpha\beta\gamma}(t) \sum_{m_i,n_i,k_i,m,n} m\times n \prod_{i=1}^{\alpha} x_{m_i} \prod_{i=1}^{\beta} y_{n_i} \prod_{i=1}^{\gamma} z_{k_i} \times$$
$$\delta_{m_1+\cdots+m_\alpha+n_1+\cdots+n_\beta+k_1+\cdots+k_\gamma,0} \qquad (69.2)$$

چون برای مدهای پایین داریم:

$$Str(T_{m_1}T_{m_2}\cdots T_{m_N}) = \delta_{m_1+m_2+\cdots+m_N,0} + O(\frac{1}{N^2}) \qquad (70.2)$$

می‌توان نوشت

$$Str([X,Y]X^\alpha Y^\beta Z^\gamma) = i\sum m\times n x_m y_n \delta_{m_1+\cdots+m_\alpha+n_1+\cdots+n_\beta+k_1+\cdots+k_\gamma,0} \prod_{i=1}^{\alpha} x_{m_i} \prod_{i=1}^{\beta} y_{n_i} \prod_{i=1}^{\gamma} z_{k_i}$$
$$(71.2)$$

و

$$Str([X,Y]\sum_{\alpha\beta\gamma}A^{\alpha\beta\gamma}(t)X^\alpha Y^\beta Z^\gamma) =$$
$$i\sum_{\alpha\beta\gamma}A^{\alpha\beta\gamma}(t)\sum m\times n x_m y_n \delta_{m_1+\cdots+m_\alpha+n_1+\cdots+n_\beta+k_1+\cdots+k_\gamma,0}\prod_{i=1}^{\alpha}x_{m_i}\prod_{i=1}^{\beta}y_{n_i}\prod_{i=1}^{\gamma}z_{k_i}$$
(۷۲.۲)

با تعریف $A_{+xy} = \sum A^{\alpha\beta\gamma}(t)X^\alpha Y^\beta Z^\gamma$ و با استفاده از (۷۲.۲) و (۲۴.۲) مدل ماتریسی غشا در پیمانه‌ی مخروط نور عبارت می‌شود از

$$H = 2\pi^2 Str(\dot{X}^2 + \dot{Y}^2 + [X,Y]^2 - 6[X,Y]A_{+xy}[X]) + O(\frac{1}{N^2}) \qquad (۷۳.۲)$$

اگر مولفه‌هایی دیگر از پتانسیل سه‌فرمی صفر نباشند، هامیلتونی عبارت می‌شود از

$$\begin{aligned}H &= \frac{1}{2}\int d^2\sigma\left((P_x+6\{x,y\}A_{xxy})^2 + (P_y+6\{x,y\}A_{yxy})^2 + \{x,y\}^2\right)\\ &+ 6\int d^2\sigma\{x,y\}A_{+xy}\end{aligned} \qquad (۷۴.۲)$$

برای جمله‌های جدیدی که هامیلتونی ظاهر شده‌اند نیز می‌توان فرایند ماتریسی را تکرار کرد و به دست آورد:

$$\begin{aligned}H &= 2\pi^2 Srt\left((P_x+6[X,Y]A_{xxy}[X])^2 + (P_y+[X,Y]A_{yxy})^2 +\right.\\ &\left.+\frac{4}{\nu^2}[X,Y]^2 + 6[X,Y]A_{+xy}\right) + O(\frac{1}{N^2})\end{aligned} \qquad (۷۵.۲)$$

که در این رابطه پتانسیل سه‌فرمی تبدیل به پتاسیل سه‌فرمی برای ماتریس‌ها شده است. دقت کنید که در رد-متقارن باید پتانسیل سه‌فرمی را بسط تیلور دهید، و رد متقارن را بر جملات بسط تیلور و $[X,Y]$ اعمال کنید.

با استفاده از (۷۰.۲) می‌توان دید که شرط گاوس تبدیل می‌شود به

$$[P_x + 6Sym\{[X,Y]A_{xxy}[X]\}, X] + [P_y + 6Sym\{[X,Y]A_{yxy}[X]\}, Y] \approx 0 \qquad (۷۶.۲)$$

که در آن منظور از نماد $Sym$ این است که پتانسیل سه‌فرمی ماتریسی را حول ماتریس صفر بسط داده شود و حاصل‌ضرب این بسط و $[X,Y]$ نسبت به تمام جایگشت‌های ممکن متقارن شود. لازم به ذکر است که گویا استفاده از رد متقارن توسط Tseytlin در [۱۶] در مدل ناجابه‌جایی برن این‌فلد ارایه شده است. مرسوم است که ماتریس‌های $X$ و $Y$ و $Z$ را با یک باز مقیاس (۵۱.۲) تعریف کنیم و از مدل ماتریسی زیر استفاده کنیم

$$H = \frac{1}{2} Srt \left( (P_x + 6[X,Y]A_{xxy}[X])^2 + (P_y + [X,Y]A_{yxy})^2 + \right.$$
$$\left. +[X,Y]^2 + 6[X,Y]A_{+xy} \right) + O(\frac{1}{N^2}) \qquad (۷۷.۲)$$

که این هنجارش نیاز به این دارد تعریف پتانسیل سه‌فرم را هم بر اساس این باز تعریف تغییر دهیم. دقت کنید که استفاده از رد متقارن هامیلتونی مدل غشا در پیمانه‌ی نور را با دقت $\frac{1}{N^2}$ به ما ارایه می‌دهد.

## ۵.۲   مدل ماتریسی برای غشا دلخواه

گسترش دستاوردهای بخش پیشین برای یک غشا با توپولوژی دلخواه در فضای هدف دلخواه ساده می‌باشد. برای یافتن مدل ماتریسی برای غشا دلخواهتان در فضای هدف دلخواه در حضور میدان‌های دلخواه از نسخه‌ی زیر استفاده کنید:

۱. پارامتربندی‌ غشا برای نوسان‌های هموار با پارامترهای مناسب.

۲. نوشتن هامیلتونی غشا در پیمانه‌ی مخروط نور.

۳. تعریف مختصه‌های ماتریسی $X^1, X^2, \cdots, X^D$ با کمک ماتریس‌های تعمیم‌یافته‌ی $T_m$ و مولفه‌های تبدیل مناسب $x^1, x^2, \cdots, x^D$. دقت کنید در غشا چنبره‌ای تبدیل مناسب عبارت است از تبدیل فوریه. در غشای

۲۹

کروی تبدیل مناسب نوشتن بر حسب هماهنگهای کروی است. برای یک غشای دلخواه باید پایه‌های مناسب که تمام نوسانات غشا را بر حسب آن بتوان بسط داد بیابید.

۴. براکت پواسون را به جابه‌جاگر تبدیل کنید.

۵. تعریف توابع ماتریسی با جای‌گزاری مختصه‌های ماتریسی به جای مختصه‌های عددی در آن ماتریس‌ها.

۶. عوض کردنِ $\int d^2\sigma$ با ردِ متقارن

۷. استفاده از ماتریس‌های $T$ در ربط دادن مختصه‌های عددی به مختصه‌های ماتریسی. اعتقاد بر این است که مستقل از توپولوژی‌ی غشا این کار را همیشه می‌توان انجام داد [۲۳]. ما این کار را برای غشای چنبره‌ای انجام دادیم. مرجع [۸] این کار را برای غشا کروی انجام داده است. این کار باید برای هر غشایی ممکن باشد. مدل ماتریسی را برای یک غشا باز با شرایط مرزی دلخواه بنویسید و چند تا از غشاها را به هم بچسبانید تا غشا مورد نظرتان به دست بیاید.

۸. در شرط گاوس هر جا که ایهامی در مورد تعریف تابع‌های ماتریسی وجود دارد از نسخه‌ی متقارن‌سازی استفاده شود.

## ۶.۲   نگاهی به ناپایداری غشا و رشد مو بر آن در مدل ماتریسی

هنجارش ماتریسی که در این فصل مرور کردیم و آن را گسترش دادیم یک فرایند کاملا کلاسیکی می‌باشد. ابتدا به سکون هیچ دلیلی نداریم که آیا مکانیک کوانتمی‌ی مدل ماتریسی با مکانیک کوانتمی‌ی غشا یکسان است یا نه؟ در این بخش می‌خواهیم یکی از نشانه‌های یکسان بودن مکانیک کوانتمی‌ی غشا و مکانیک کوانتمی‌ی مدل ماتریسی را بررسی کنیم.

در انتهای فصل قبل نشان دادیم که مکانیک کوانتمی‌ی غشا باید معضل رشد مو بر غشا کلاسیک را حل کند. در این بخش قصد داریم ناپایداری‌ی رشد مو بر غشا را در مدل ماتریسی بیابیم و واکاوی کنیم که مکانیک کوانتمی‌ی مدل ماتریسی آیا و چگونه این ناپایداری را رفع می‌کند. این یک نشانه‌ی یکسان بودن مکانیک کوانتمی‌ی غشا و مدل ماتریسی است.

در هنجارش ماتریسی غشا، ناپایداری‌ی رشد مو خود را به صورت مسیرهای نامحدود که پتانسیل بر روی آنها صفر

می‌شود نشان می‌دهد. به عنوان مثال در صورت وجود دو ماتریس با درآیه‌های نا صفر به صورت

$$X^1 = \begin{pmatrix} x & 0 \\ 0 & 0 \end{pmatrix}, \ X^2 = \begin{pmatrix} 0 & y \\ y & 0 \end{pmatrix}, \tag{2.78}$$

پتانسیل $tr([X^1, X^2]^2)$ عبارت می‌شود از $V(x,y) = x^2 y^2$. همان‌طور که از شکل پتانسیل آشکار است پتانسیل بر روی محورهای $x$، $y$ صفر است و به طور کلاسیکی $x$ و $y$ می‌توانند تا بی‌نهایت بروند. در مکانیک کوانتمی باید از تابع موج $\Psi(x,y)$ که $|\Psi(x_0, y_0)|^2$ احتمال دیده شدن $x_0$ و $y_0$ به ترتیب برای متغیرهای $x$ و $y$ استفاده کرد. معادله‌ی شرودینگر تابع موج عبارت می‌شود از

$$-\frac{\hbar}{2m}(\frac{\partial^2}{\partial x^2} + \frac{\partial^2}{\partial y^2})\Psi(x,y) + x^2 y^2 \Psi(x,y) = E\Psi(x,y) \tag{2.79}$$

که در رابطه‌ی بالا $m$ جرم $D$-ذره است. ما علاقه‌مند به بررسی رفتار $\Psi(x,y)$ برای $x$ها و $y$های بزرگ در راستای محورهای مختصات می‌باشیم. پس بدون از دست دادن کلیت مساله خود را محدود به بررسی این رفتار بر روی محور $x$ می‌کنیم. در فاصله‌ی دور از مبدا در همسایگی نقطه‌ی $(x_0, 0)$ می‌توان معادله‌ی موج را به صورت زیر تقریب زد

$$-\frac{\hbar}{2m}(\frac{\partial^2}{\partial x^2} + \frac{\partial^2}{\partial y^2})\Psi(x,y) + x_0^2 y^2 \Psi(x,y) = E\Psi(x,y) \tag{2.80}$$

که معادله‌ی موج برای نوسانگر هماهنگ ساده می‌باشد و ترازهای انرژی آن عبارتند از

$$E_n = \hbar\sqrt{\frac{2}{m}}|x_0|(n + \frac{1}{2}) \tag{2.81}$$

و کمینه‌ی انرژی عبارت می‌شود از

$$E_0 = \frac{\hbar|x_0|}{\sqrt{2m}} \tag{2.82}$$

پس اگر $D$-ذره‌ای را با هر انرژی محدودی در مبدا قرار دهیم بر خلاف انتظاری که در مکانیک کلاسیک داشتیم، ذره نمی‌تواند در راستاهای مختلف تا بی‌نهایت دور برود. بیشینه فاصله‌ای که ذره می‌تواند از مبدا دور شود عبارت است از

$$l = \frac{E\sqrt{2m}}{\hbar}. \qquad (2.83)$$

این دقیقن همان اتفاقی است که در مورد رشد مو در سطح غشا در مکانیک کوانتمی اتفاق می‌افتد. غشای کوانتمی موی بلند ندارد و تقریبن کچل است. در مکانیک کوانتمی مدل ماتریسی هم $D$-ذره‌ها در بند می‌شوند و نمی‌توانند از مبدا دور شوند. با در نظر گرفتن ابر تقارن ابهامی پیش می‌آید که نگارنده جوابش را نمی‌داند. در مدلهای ابرتقارنی انرژی حالت پایه صفر می‌ماند و ذرات می‌توانند در مکانیک کوانتمی مدل ماتریسی تا بی‌نهایت بروند. اما به خاطر اصل عدم قطعیت هایزنبرگ -همانگونه که در پایان فصل قبل به آن اشاره کردیم- موهای غشای ابرمتقارن نمی‌توانند تا بی‌نهایت رشد کنند. مرجع [۱۱] به این ابهام توجه نمی‌کند. رفع این ابهام نیازمند به بررسی و واکاوی‌های بیشتر دارد.

# فصل ۳

# کنش برن-اینفلد و تعمیم آن برای $D$-غشاها در نظریه‌های پیمانه‌ای

## ۳.۱ کنش برن-اینفلد

میدان الکتریکی‌ای که قانون کولمب بیان می‌کند در فاصله‌ی نزدیک به تک بار نقطه‌ای به صورت $\frac{1}{r^2}$ که $r$ فاصله از بار است واگرا می‌شود. واگرایی میدان الکتریکی از دیدِ فیزیکی قابل قبول نیست و باید به گونه‌ای نظریه‌ی الکترومغناطیسی‌ی کلومبی تصحیح شود. مکانیک کوانتمی این کار را انجام می‌دهد. در مکانیک کوانتمی وقتی به بار نزدیک می‌شویم به خاطر اصل عدم قطعییت هایزنبرگ یک توزیع بار می بینیم و میدان الکتریکی دیگر بی‌نهایت نیست. راه حل دیگر برای رفع این مشکل بی‌نهایت شدن٬ تغییر دادن قانون کلومب در فاصله‌های بسیار نزدیک به بار هست. Mei در سال ۱۹۱۲ میلادی پیشنهاد داده است که میدان الکتریکی نمی‌تواند از یک کران بالا بزرگتر شود. او پیشنهاد بازتعریف میدان الکتریکی را به صورت زیر می‌دهد [۱]

$$E_{eff} = \frac{1}{\sqrt{1 - \frac{E^2}{E_0^2}}} \ , \ E \propto \frac{1}{r^2}$$

در این مدل میدان الکتریکی ($E_{eff}$) ناتکین و خودانرژی‌ی تک بار نقطه‌ای محدود می باشد و میدان الکتریکی در فاصله‌های دور به صورت $r^{-2}$ افت پیدا می‌کند. اما این جواب نسبت به تبدیلات لورنتس به صورت هموردا به دست نیامده است. در سال ۱۹۳۲ و ۱۹۳۴ میلادی برن-اینفلد یک مدل غیر خطی برای الکترودینامیک ارایه دادند که چگالی‌ی لاگرانژی عبارت

۳۳

است از

$$L_{BI} = \beta^2 \sqrt{\det(\delta^\mu_\nu + \beta^{-2} F^\mu_\nu)} \qquad (1.3)$$

که $\beta$ پارامتری ثابت است و نقش میدانی حدی مدل Mei را بازی می‌کند. چگالی لاگرانژی بالا را در چار بعد می‌توان بر حسب کمیتهای ناوردای لورنتس زیر

$$P = \frac{1}{4} F_{\mu\nu} F^{\mu\nu} \, , \, S = \frac{1}{4} \epsilon^{\mu\nu\lambda\rho} F_{\mu\nu} F_{\lambda\rho}$$

به صورت زیر نوشت

$$L_{BI} = \beta^2 \left(1 - \sqrt{1 + \frac{2P}{\beta^2} - \frac{S^2}{\beta^4}}\right) \qquad (2.3)$$

یا

$$L_{BI} = \beta^2 \left(1 - \sqrt{1 + \frac{1}{2\beta^2}(B^2 - E^2) - \frac{1}{16\beta^4}(E.B)^2}\right) \qquad (3.3)$$

که در آن $E$ و $B$ میدانهای الکتریکی و مغناطیسی هستند. معادلات بالا نشان می‌دهند که کران بالای $\sqrt{2}\beta$ برای میدان الکتریکی ناشی از یک تک بار ساکن وجود دارد. برای سریع دیدن این موضوع کافی است کنش (3.3) را برای میدان الکتریکی بنویسیم

$$L_{BI} = \beta^2 \left(1 - \sqrt{1 - \frac{E^2}{2\beta^2}}\right) \qquad (4.3)$$

که بر حسب پتانسیل الکتریکی تبدیل می‌شود به

$$L_{BI} = \beta^2 \left(1 - \sqrt{1 - \frac{(\nabla\Phi)^2}{2\beta^2}}\right) \, , \, -\nabla\Phi = E \qquad (5.3)$$

معادله‌ی حرکت $\Phi$ عبارت می‌شود از

$$\partial_\mu \left( \frac{\partial^\mu \Phi}{\sqrt{1 - \frac{(\nabla \Phi)^2}{2\beta^2}}} \right) = 0 \qquad (3.6)$$

برای پتانسیل ایستای $\Phi(x,t) = \Phi(x)$ می‌توان دید که $-\nabla \Phi = \frac{\hat{r}}{\sqrt{r^4 + \frac{1}{2\beta^2}}}$ همه‌جا به جز مبدا در معادله‌ی حرکت صدق می‌کند:

$$\partial_\mu \left( \frac{-\partial^\mu \Phi}{\sqrt{1 - \frac{(\nabla \Phi)^2}{2\beta^2}}} \right) = 4\pi \delta(\vec{r}) \qquad (3.7)$$

پس حل $E \propto \frac{1}{\sqrt{r^4 + \frac{1}{2\beta^2}}}$ میدان الکتریکی‌ی یک تک بار ساکن را در نظریه‌ی الکترومغناطیس برن-اینفلد ارایه می‌دهد. در فاصله‌های دور از این بار این میدان تبدیل به میدان الکتریکی‌ی معمولی می‌شود

$$E \approx \frac{\vec{r}}{r^3} \qquad (3.8)$$

اما میدان الکتریکی در روی بار بی‌نهایت نیست و خود انرژی‌ی تک بار نقطه‌ای در نظریه‌ی برن-اینفلد محدود است:

$$\int E^2 d^3 r = 2\pi^2 \beta \qquad (3.9)$$

کنش برن-اینفلد هم‌چنین دارای خصوصیت‌های زیبای دیگر از جمله خاصیت‌های زیر است:

1. در این کنش موج-ضربه وجود ندارد و از هر حالت اولیه‌ی ناتکینی که برای میدانها شروع کنیم هیچ‌گاه به حالتی نمی‌رسیم که کمیتی فیزیکی تکین شود [18].

2. پدیده‌ی دوشکستی در این نظریه روی نمی‌دهد. قطبش‌های مختلف نور با سرعت یکسان حرکت می‌کنند [20].

3. یگانه کنشی است که برای میدانهای ضعیف هم‌چون کنش معمولی‌ی الکترودینامیک می‌شود و دو خاصیت پیش را

۳۵

داراست [۱۹].

## ۳.۲ کنش برن-اینفلد و چرن-سایمون برای تک $D$ غشا

$D_p$ غشا موجودی $p+1$ بعدی در فضا-زمان می‌باشد که سرهای ریسمان باز مقید به حرکت آزادانه بر روی آن می‌باشند. مدهای بی‌جرم یک ریسمان باز که هر دو سر آن روی یک غشا می‌باشند یک نظریه‌ی پیمانه‌ای $U(1)$ با بردار $A_a$، $a = 1, \cdots, p+1$ و $9 - p$ میدان نرده‌ای حقیقی $\phi^i$، $i = p+1, \cdots, 9$ و شریک‌های ابرمتقارنشان می‌سازد.

انرژی پایین $D_p$ غشا را می‌توان با کمک نظریه‌ی ابریانگ-میلز در ده بعد که به $p+1$ یک بعد کاهش یافته است توصیف نمود. Leigh با در نظر گرفتن تمام تصحیحات بالاتر نظریه‌ی ریسمان برای میدان‌های ثابت نشان داده است که کنش مربوط به غشا به صورت گسترشی از کنش برن-اینفلد به صورت زیر است [۲۴] [1]

$$S_{BI} = -T_p \int d^{p+1}\sigma\, e^{-\phi} \sqrt{-\det(P[G+B]_{ab} + 2\pi l_s^2 F_{ab})} \qquad (3.10)$$

که $T_p$ انرژی بر واحد حجم غشا می‌باشد. تمام تصحیح‌های مرتبه‌های بالاتر مادامی قابل اعتماد می‌باشند که تغییرات میدان‌ها در فاصله‌ی $l_s$ کوچک باشد. این کنش همچنین نشان می‌دهد که غشاها موجوداتی متحرک می‌باشند که جابه‌جایی آنها با $\Phi^i$ توصیف می‌شود $\Delta X^i = 2\pi l_s^2 \Phi^i$.

رابطه‌ی (3.10) تنها جفت شدن غشا را به مدهای بی‌جرم بخش نِو-شوارتز زمینه‌ی ریسمان بسته نشان می‌دهد. غشاها می‌توانند با پتانسیل‌های فرمی بخش رامُند-رامُند هم جفت شوند [۲۵]. این برهم‌کنش‌ها در کنش چِرن-سایمون بیان می‌شوند

$$S_{Sc} = \mu_p \int P[\sum_n C^{(n)} \wedge e^B] \wedge e^{2\pi l_s^2 F} \qquad (3.11)$$

که $C^{(n)}$ پتانسیل رامُند-رامُند $n+1$ فرمی و $\mu_p$ بار رامُند-رامُند غشا هست. منظور از $C^{(n)} \wedge e^B$ ضرب خارجی

---

[1] ما در این پایان‌نامه نظریه‌ی ریسمان را مرور نمی‌کنیم. اما تمام اطلاعاتی که خواننده‌ی ناآشنا به موضوع لازم دارد در مراجع [۱۳، ۱۴، ۱۵، ۱۶، ۱۷، ۱۸، ۱۹، ۲۰، ۲۱، ۲۲، ۲۳، ۲۴، ۲۵، ۲۶، ۲۷، ۲۸، ۲۹، ۳۰، ۳۱] وجود دارند.

زیر می‌باشد

$$C^{(n)} \wedge e^B = C^{(n)} + C^{(n)} \wedge e^B + \frac{1}{2} C^{(n)} \wedge B \wedge B + \cdots$$

می‌باشد. $e^{2\phi l_s^2 F}$ نیز همچون $e^B$ تعریف می‌شود. انتگرال‌گیری تنها بر روی تانسورهای هم‌مرتبه با بعد جهان‌حجم $D_p$ غشا انجام می‌شود. این کنش به این معنی است که یک غشا، در حضور میدانهای دیگر با پتانسیلهای راموند-راموند مرتبه‌ی پایین‌تر را نیز حمل می‌کند [26] برای این که موضوع را بهتر ببینیم اجازه دهید کنش یک تک $D_2$ غشا را در حضور میدان $F$ و پتانسیل یک-فرم $C^{(1)}$ و نبود میدانهای دیگر با جزییات بررسی کنیم. در این حالت کنش چرن-سایمون عبارت می‌شود از:

$$S_{CS} = 2\pi l_s^2 \mu_2 \int C_\mu^{(1)} \frac{\partial x^\mu}{\partial \sigma^a} F_{bc} \epsilon^{abc} d\sigma^0 d\sigma^1 d\sigma^2 \qquad (12.3)$$

اگر تنها مولفه‌ی $F_{12}$ از $F_{ab}$ غیر صفر باشد آنگاه (12.3) به صورت زیر ساده خواهد شد

$$F_{12} = \epsilon_{12} F \qquad (13.3)$$

$$S_{CS} = 4\pi l_s^2 \mu_2 \int d\sigma^0 \int P[C^{(1)}]_0 F d\sigma^1 d\sigma^2 \qquad (14.3)$$

این کنش، کنش یک سطح دو بعدی با توزیع بار سطحی مربوط به پتانسیل تک فرم با چگالی سطحی $4\pi l_s^2 \mu_2 F$ را در حضور میدان $C^{(1)}$ می‌باشد. به عبارتی دیگر در حضور میدانِ $F$ غشا گویا تعدادی $D_0$ غشا بر روی $D_2$ غشا قرار گرفته‌اند و یک حالت در هم تنیده و مقید تشکیل داده‌اند. این گونه پیکربندی‌ها برای اولین بار در [28] به صورت حالت مقید غشاها با بعدهایی مختلف تعبیر شد. ما این موضوع را در فصل بعد با جزییات بررسی می‌کنیم.

## ۳.۳ کنش برن-اینفلد و چرن-سایمون برای غشاهای روی هم‌افتاده

یک خاصیت مهم $D$-غشاها این است که هنگامی که $N$ تا از آنها را دقیقا روی هم می‌گذاریم تقارن آنها به تقارن ناجابه‌جایی $U(N)$ افزایش پیدا می‌کند [۲۸]. وقتی غشاها به هم نزدیک می‌شوند جرم مد پایه‌ی ریسمان‌هایی که بین این غشاها چسبیده‌اند صفر می‌شود و این مدهای بی‌جرم تقارن را افزایش می‌دهند [۲۹]. Myers با استفاده از دوگان-T نشان داده است که با تقریب خوبی کنش برن-اینفلد برای چند غشای روی هم افتاده عبارت است از [۱۰]:

$$S_{NBI} = -T_p \int d^{p+1}\sigma \, Str\left(e^{-\phi}\sqrt{-\det(P[E_{ab} + E_{ai}(Q^{-1} - \delta)^{ij}E_{jb}] + 2\pi l_s^2 F_{ab})\det(Q^i_j)}\right)$$

(۳.۱۵)

که در آن از نمادهای $a, b, c, \cdots$ برای راستاهای موازی با جهان-حجم و از نمادهای $i, j, k, \cdots$ بر راستاهای عمود بر جهان-حجم استفاده شده است و

$$E = G + B \tag{۳.۱۶}$$

$$Q^i_j = \delta^i_j + 2\pi i l_s^2 [\Phi^i, \Phi^k]E_{kj} \tag{۳.۱۷}$$

$$(Q^{-1} - \delta)^{ij} = (Q^{-1} - \delta)^i_k E^{kj} \, , \, E^{ij}E_{jk} = \delta^i_k \tag{۳.۱۸}$$

همچنین در حضور میدانهای پیمانه‌ای تمام مشتق‌گیرها، مشتق‌گیری هم‌وردا می‌باشند: $D_a\Phi^i = \partial_a\Phi^i + i[A_a, \Phi^i]$ [۱۲]. رد متقارن در مرتبه‌ی $F^6$ نیاز به تصحیح دارد [۳۰]. کنش چرن-سایمون برای جهان‌حجم ناجابه‌جایی به صورت زیر می‌باشد [۱۰]:

$$S_{NCS} = \mu_p \int Str\left(P[e^{2\pi l_s^2 i_\Phi i_\Phi}\sum C^{(n)}e^B]e^{2\pi l_s^2 F}\right) \tag{۳.۱۹}$$

در کنش بالا علاوه بر گسترش میدانها به میدانهای ناجابه‌جایی با تبدیل پارامترهای $x$ به ماتریسهای $X$ و استفاده از رد متقارن، عمل‌گری نیز درون بر‌ان اضافه شده است. $i_\Phi$ ضرب داخلی با $\Phi$ می‌باشد. ضرب داخلی یک عمل با درجه‌ی فرمی منفی

۳۸

یک میباشد. ضرب داخلی را با مثال پایین برای خوانندهی ناآشنا با این مفهوم مرور میکنیم:

$$C^{(2)} = \frac{1}{2} C^{(2)}_{\mu\nu} dx^\mu dx^\nu$$

$$i_v C^{(2)} = v^\mu C^{(2)}_{\mu\nu} dx^\nu$$

$$i_v i_w C^{(2)} = w^\nu v^\mu C^{(2)}_{\mu\nu} = -i_w i_v C^{(2)} \qquad (3.20)$$

برای بردارهای معمولی داریم

$$(i_v)^2 = 0 \qquad (3.21)$$

اما در مدلهای ناجابهجایی که پارامترهای عددی به ماتریس تبدیل شدهاند داریم

$$i_\Phi i_\Phi C^{(2)} = \Phi^j \Phi^i C^{(2)}_{ij} = \frac{1}{2}[\Phi^i, \Phi^j] C^{(2)}_{ij} \qquad (3.22)$$

بنابراین حضور $e^{2\pi l_s^2 i_\Phi i_\Phi}$ در (3.19) یک تصحیح نابدیهی به کنش میگذارد. پس در تعمیم ناجابهجایی کنش تک غشا D میبینیم که حالت مقید غشاها میتواند به پتانسیلهای راموند-راموند بالاتر جفت شود. برای این که این جفت شدگی را خوب ببینیم اجازه دهید کنش $D_0$ غشا را وقتی $F$ صفر است بنویسیم

$$S_{NCS} = \mu_0 \int dt Str \left( C^{(1)}_t + 2\pi l_s^2 C^{(1)}_i D_t \Phi^i + \pi i l_s^2 (C^{(3)}_{ijk}[\Phi^k, \Phi^j] + 2\pi l_s^2 C^{(3)}_{ijkl} D_t \Phi^i [\Phi^k, \Phi^j]) + \cdots \right)$$

$$(3.23)$$

لازم به ذکر است که بار و کشش سطحی غشاها با هم برابر هستند [21]

$$\mu_0 = T_0 \qquad (3.24)$$

(3.23) جفت شدن یک $D_0$ غشا را با پتانسیلهای فرمی بالاتر از یک را به نمایش میگذارد.

## ۴.۳  حالت مقید $D_0$ غشاها و هنجارش ماتریسی‌ی غشا

برای $D_0$ غشاهای روی هم‌افتاده یک جفت شدگی با $C^{(3)}$ که پتانسیل مربوط به $D_2$ غشا است وجود دارد:

$$۲\int Str(P[i_\Phi i_\Phi C^{(3)}]) = \int dt Str(C^{(3)}_{tjk}[\Phi^k, \Phi^j] + ۲\pi l_s^۲ C^{(3)}_{ijk} D_t \Phi^k [\Phi^i, \Phi^j]) \qquad (۲۵.۳)$$

اگر حد $N \to \infty$ را در نظر بگیریم رابطه‌ی بالا دقیقا همان کنشی است که در هنجارش ماتریسی‌ی یک غشا در حضور پتانسیل سه-فرم دل‌خواه به دست آوردیم. این اثباتی بر این مطلب است که هنجارش ماتریسی‌ای‌که ارایه دادیم درست است.

این شباهت را می‌توان این گونه توصیف کرد که تعداد زیادی $D_0$ غشا یک $D_2$ غشا را ساخته‌اند.

به سبب وجود جفتیدگی با پتانسیل‌های فرمی بالاتر در (۲۳.۳) می‌توان انتظار داشت که $D$ غشاهایی با هر بعد دل‌خواهی را بتوان به صورت حالت مقید $D_0$ غشاها توصیف کرد. اما چنین توصیفی یا مثالی از آن هنوز در جایی گزارش نشده است.

# فصل ۴

# حالت مقید D ذره‌ها

از این لحظه به بعد به $D_\circ$ غشا یک $D$-ذره می‌گوییم. در بخش‌های پیش مرور کردیم که انرژی‌های پایین $N$ $D$-ذره در نظریه‌ی میدان‌های پیمانه‌ای با تبدیل پارامترهای عددی‌ی فضا به ماتریس‌های $N \times N$ یکانی با کمک کنش زیر توصیف می‌شود

$$
\begin{aligned}
S_{NBI} &= -T_\circ \int d\tau\, Str(e^{-\phi}\sqrt{\det P[E_{\circ\circ} + E_{\circ i}(Q^{-1} - \delta)^{ij}E_{j\circ}]\det Q^i{}_j}) + \\
&\quad + \mu_\circ \int Srt(P[e^{i\lambda i_\Phi i_\Phi}\sum C^{(n)}e^B])
\end{aligned}
\tag{۴.۱}
$$

$$Q^i{}_j = \delta^i{}_j + i\lambda[\Phi^i, \Phi^j]E_{kj} \tag{۴.۲}$$

$$E = G + B \tag{۴.۳}$$

$$\lambda = ۲\pi l_s^۲ \tag{۴.۴}$$

که تمام میدان‌ها در حالت کلی تابعی از مختصه‌های ماتریسی‌ی $\Phi^i$ ها می‌باشند. شکل این کنش پیچیده است و کار کردن با آن ساده نیست. در حالتی که تنها دو $D$ذره به هم تنیده باشند می‌تواند در موردهای ساده‌ای کنش را به طور صریح بر حسب توابع بیضوی نوشت [۱۷]. اما ما در اینجا علاقه‌مند به واکاوی حالت‌های مقید تعداد زیادی $D$-ذره هستیم. پس روندی که توسط [۱۰، ۱۳] ارایه شده را برمی‌گزینیم و گسترش می‌دهیم.

۴۱

پیش از آن که جلوتر برویم ابتدا باید بفهمیم که اگر $N$ $D$-ذره از کجا باید داشته باشیم بفهمیم آنها در هم تنیده می‌باشند و تنها یک حالت مقید داریم نه مجموع چند حالت مقید جدا از هم. اگر بتوان تمام مختصات ماتریسی $\Phi^i$ ها را با هم بلوک قطری کرد آنگاه ما چند حالت مقید جدا از هم داریم. وقتی که نمی‌توانیم این کار را بکنیم یک تک حالت مقید داریم.

## ۴.۱  کره‌ی مه‌گون

منظور ما در این پایان‌نامه از مه‌گون، fuzzy می‌باشد. در فضا-زمان تخت و نبود میدانهای E و B با انتخاب $t = \sigma^\circ$ و پیمانه‌ی ایستا کنش برن-اینفلد برای N $D$-ذره عبارت می‌شود از

$$S_{BI} = -T_\circ \int dt\, Str\sqrt{(۱ - \lambda^۲ D_t\Phi^i Q_{ij}^{-۱} D_t\Phi^j)\det Q_{ij}} \qquad (۵.۴)$$

$$Q_{ij} = \delta_{ij} + i\lambda[\Phi^i, \Phi^j] \qquad (۶.۴)$$

که بسط آن تا مرتبه‌ی دوم $\lambda^۲$ عبارت است از

$$S_{BI} = \int dt(T - V) \qquad (۷.۴)$$

$$T = \frac{T_\circ \lambda^۲}{۲} Tr(D_t\Phi^i D_t\Phi^i) \qquad (۸.۴)$$

$$V = -\frac{T_\circ \lambda^۲}{۴} tr([\Phi^i, \Phi^j]^۲) \qquad (۹.۴)$$

در حضور پتانسیل سه‌فرم کنش چرن-سایمون برای $D$-ذره‌ها عبارت می‌شود

$$S_{CS} = i\lambda\mu_\circ \int Str P[i_\Phi i_\Phi C^{(۳)}]. \qquad (۱۰.۴)$$

در حالتی که پتانسیل سه‌فرم تابعی خطی از $\Phi$ باشد این کنش به صورت زیر ساده می‌شود

$$S_{CS} = i\lambda^2\mu_\circ \int dt\, Str[\Phi^i\Phi^j(\Phi^k\partial_k C^{(3)}_{ijk}(t) + C^{(3)}_{ijk}D_t\Phi^k)] \tag{۴.۱۱}$$

$$= \frac{i}{٣}\lambda^2\mu^\circ \int dt\, tr(\Phi^i\Phi^j\Phi^k)F^{(4)}_{tijk}(t) \tag{۴.۱۲}$$

که $F^{(4)}_{tijk}(t)$ قدرت میدان پتانسیل سه‌فرمی است. برای رسیدن به سطر دوم عبارت بالا از انتگرال‌گیری جز‌به‌جز استفاده کرده‌ایم. در حالتی که قدرت میدان پتانسیل سه‌فرمی به صورت زیر باشد

$$F^{(4)}_{tijk}(t) = \begin{cases} -٢f\epsilon_{ijk} & i,j,k \in \{١,٢,٣\} \\ ٠ & otherwise \end{cases} \tag{۴.۱۳}$$

پتانسیل ناشی از $S_{SC}$ و $S_{BI}$ عبارت می‌شود از

$$V(\Phi) = -\frac{\lambda^2 T_\circ}{۴}tr([\Phi^i,\Phi^j]^٢) + \frac{٢i\lambda^2}{٣}\mu_\circ tr(\Phi^i\Phi^j\Phi^k)\epsilon_{ikj}f \tag{۴.۱۴}$$

وردش اول پتانسیل نسبت به $\Phi$ معادله‌ی فرین‌های پتانسیل را به ما می‌دهد

$$[[\Phi^i,\Phi^j],\Phi^j] + if_{ijk}[\Phi^j,\Phi^k] = ٠ \tag{۴.۱۵}$$

ماتریس‌های جابه‌جا شونده‌ی $\Phi^i$ در معادله‌ی بالا صدق می‌کنند و برای آنها $V(\Phi) = ٠$ می‌باشد. به نام یک حدس برای جواب ناجابه‌جایی اجازه دهید فرض کنیم $\Phi^i$ در رابطه‌ی زیر صدق می‌کنند:

$$[\Phi^i,\Phi^j] = ٢iR\epsilon_{ijk}\Phi^k, \; i,j,k \in \{١,٢,٣\} \tag{۴.۱۶}$$

که $R$ عددی ثابت می‌باشد. با جای‌گذاری (۱۶.۴) در (۱۵.۴) می‌بینیم که برای مقدار زیر رابطه‌ی (۱۵.۴) برآورده می‌شود:

$$R = \frac{f}{۲}. \qquad (۱۷.۴)$$

بنابراین یک جواب نابدیهی معادله‌ی (۱۵.۴) عبارت است

$$\Phi^i = \frac{f}{۲}\alpha^i \qquad (۱۸.۴)$$

که $\alpha^i$ ها می‌توانند هر نمایش یکانی $N \times N$ جبر $SU(۲)$ باشند:

$$[\alpha^i, \alpha^j] = ۲i\epsilon_{ijk}\alpha^k \qquad (۱۹.۴)$$

در ابتدا خود را محدود به بررسی نمایش یکانی کاهش‌ناپذیر جبر $SU(۲)$ می‌کنیم که با انتخاب پایه‌های مناسب برای این جبر داریم

$$tr[\alpha^i, \alpha^j] = \frac{N}{۳}(N^۲ - ۱)\delta^{ij} \qquad (۲۰.۴)$$

و پتانسیل (۱۵.۱) برای (۲۰.۴) و (۱۸.۴) ساده می‌شود به

$$V_N = -\frac{T_\circ \lambda^۲ f^۴}{۶}(N^۳ - N) \qquad (۲۱.۴)$$

و این حالت ناجابه‌جایی انرژی کمتری نسبت به حالتی که میدان‌ها با هم جابه‌جا می‌شدند دارد. پس حالت جابه‌جایی $D$-ذره‌ها کمینه‌ی مطلق انرژی سیستم نیست. $D$-ذره‌ها تمایل خواهند داشت به سمت حالت ناجابه‌جایی واهلش کنند.

نمایش کاهش‌ناپذیری که در بالا در نظر گرفتیم تنها یک جواب معادله‌ی (۱۵.۴) می‌باشد. نمایش‌های کاهش‌پذیر جبر

۴۴

$SU(2)$ نیز جواب این معادله هستند. در نمایش‌های کاهش‌پذیر، با انتخاب پایه‌های استاندارد، $tr[(\alpha^i)^2]$ نسبت به مقدارش در نمایش کاهش‌ناپذیر کوچکتر است و در نتیجه برای نمایش‌های کاهش‌پذیر داریم

$$V_N < V_r \leq 0 \qquad (22.4)$$

پس نمایش‌های نابدیهی کاهش‌پذیر مربوط به حالت‌های میانی‌ای بین حالت آزاد $D$-ذره‌ها و حالت تماما درهم‌تنیده‌ی $D$-ذره‌ها می‌باشند.

لازم به ذکر است که در معادله‌های فرین‌های (15.4) تنها جابه‌جاگر $\Phi$ ظاهر شده است. پس تحت تبدیل

$$\Phi^i \to \Phi^i + x^i I_N , \qquad (23.4)$$

پتانسیل ناوردا است. بنابراین می‌توان رد میدان‌ها را مرکز جرم $D$-ذره‌ها تعبیر کرد:

$$x^i_{cm} = \frac{1}{N} tr(\Phi^i). \qquad (24.4)$$

$\Phi^i$ ها در جبر $SU(2)$ صدق می‌کنند پس داریم

$$(\Phi^1)^2 + (\Phi^2)^2 + (\Phi^3)^2 = N(N^2 - 1)R^2 I_N \qquad (25.4)$$

اگر متغیرها عدد می‌بودند رابطه‌ی بالا یک کره را در فضای سه‌بعدی تعریف می‌کرد. اکنون که متغیرها ماتریس هستند رابطه‌ی بالا اصطلاحا یک کره‌ی مه‌گون ناجابه‌جایی را توصیف می‌کند. شعاع این کره را می‌توان به شکل زیر تعریف کرد

$$R = \lambda (\sum tr \frac{(\Phi^i)^2}{N})^{\frac{1}{2}} \qquad (26.4)$$

۴۵

شعاع ناجابه‌جایی کره‌ی مه‌گون حالت پایه (۲۱.۴) عبارت می‌شود از

$$R = \pi l_s^2 fN\sqrt{1 - \frac{1}{N^2}} \tag{۲۷.۴}$$

$$\downarrow$$

$$R_\circ = \pi l_s^2 fN , \ for\ large\ N \tag{۲۸.۴}$$

اگر جواب (۱۸.۴) را در (۱۲.۴) قرار دهیم این تکه از کنش عبارت می‌شود از

$$-R_\circ^3 (1 - \frac{1}{N^2}) \int dt F_{t123}^{(4)} \tag{۲۹.۴}$$

که می‌توان از این به عنوان قطبیدگی $D$-ذره‌ها در راستای قدرت میدان پتانسیل سه‌فرمی توصیف کرد. با جای‌گذاری (۱۸.۴) در $S_{SC} = i\lambda\mu_\circ \int Str(P[i_\Phi i_\Phi C^{(3)}])$ برهم‌کنش‌ها با ممان‌های قطبیی مرتبه‌ی بالاتر را نیز مشاهده خواهد شد.

ما برای پیدا کردن حالت پایه‌ی ناجابه‌جایی جمله‌ی $\sqrt{\det Q_{ij}}$ را در کنش برن-اینفلد تقریب زدیم و تنها چند جمله‌ی اول بسط تیلور آن را در نظر گرفتیم. چون این جمله تنها تابعی از جابه‌جاگرها می‌باشد حتی در صورت تقریب نزدن آن عبارت $\Phi^i = R\alpha_N^i$ کماکان جواب معادله‌ی حرکت خواهد بود اما $R$ دیگر از رابطه‌ی (۱۷.۴) به دست نمی‌آید. برای به دست آوردن مقدار دقیق $R$ توجه می‌کنیم که

$$\det Q_{ij} = \det(Q_{ij} - 2\lambda R^2 \epsilon_{ijk}\alpha_N^k) \tag{۳۰.۴}$$

$$= 1 + 4\lambda^2 R^4 \sum_{i=1}^{3}(\alpha_N^i)^2 - 8\lambda^3 R^6(\alpha_N^3\alpha_N^1\alpha_N^2 - \alpha_N^2\alpha_N^3\alpha_N^1) \tag{۳۱.۴}$$

در جمله‌ی آخر عبارت ابهامی وجود دارد اما به سبب رد متقارن در کنش این ابهام در مقدار کنش بی‌تاثیر است. با گرفتن

رد متقارن به دست می‌آوریم

$$Str\sqrt{\det Q_{ij}} = \sqrt{1 + 4\lambda^2 R^4(N^2-1)} \qquad (32.4)$$

با استفاده از (12.4) پتانسیل کل در حضور میدان سه‌فرمی (13.4) عبارت می‌شود از

$$V_0(R) = T_0 N \left( \sqrt{1 - (1 - \frac{1}{N^2})^{-1} \frac{4R^2}{\lambda^2 N^2}} - \frac{4}{3\lambda N} R^3 \right) \qquad (33.4)$$

برای $\frac{4R^2}{\lambda^2 N^2}$ های کوچک این عبارت به (21.4) تقلیل پیدا می‌کند.

## ۲.۴  درهم‌تنیده‌های نوین و تمام بسترهایشان

در بخش پیش حالت مقید $D$-ذره‌ها را در فضا-زمان تخت و نبودِ هیچ میدان دیگری به جز پتانسیل سه‌فرمی بررسی کردیم. دیدیم که حضور میدان سه‌فرمی $D$-ذره‌ها را قطبیده می‌کند و این $D$-ذره‌های قطبیده ابژه‌های مه‌گون پدید می‌آورند. در این بخش قصد داریم همین فرایند را در فضا-زمان خمیده و بودِ میدانهای فرمی دلخواه انجام دهیم و ببینیم هر میدانی را چگونه مه‌گونی را به $D$-ذره‌ها القا می‌کند. این کار را برای اولین بار داریم انجام می‌دهیم.

ابتدا چارچوب مرجعی انتخاب می‌کنیم که در آن سرعت مرکز جرم حالت مقید $D$-ذره‌ها صفر باشد. در این چارچوب به بررسی‌ی حالتهای درهم‌تنیده‌ی $D$-ذره‌ها در حول و حوش نقطه‌ی مرکز جرمشان می‌کنیم. مرکز جرم را با $P$ نشان می‌دهیم. در همسایگی‌ی $P$ راستاهای مسیرهای ژئودزیک عبوری از نقطه‌ی $P$ را محورهای مختصات انتخاب می‌کنیم. با این انتخاب متریک در همسایگی $P$ عبارت می‌شود از [۹]:

$$g_{\mu\nu} = \eta_{\mu\nu} \qquad (34.4)$$

$$g_{\mu\nu,\alpha} = 0 \qquad (35.4)$$

$$g_{\mu\nu} = \eta_{\mu\nu} + \frac{\lambda^2}{2}\Phi^i\Phi^j g_{\mu\nu,ij} + \frac{\lambda^3}{6}\Phi^i\Phi^j\Phi^k g_{\mu\nu,ijk} \qquad (36.4)$$

۴۷

میدانهای پیمانه‌ای $B$ و $C^{(n)}$ را به گونه‌ای تثبیت می‌کنیم که در نقطه‌ی $P$ داشته باشیم

$$B_{\mu\nu}|_P = 0 \qquad (37.4)$$

$$C^{(n)}|_P = 0 \qquad (38.4)$$

در این پیمانه بسط تیلور این میدانها بر حسب $\Phi$ ها به صورت زیر خواهد شد

$$B_{\mu\nu} = \lambda B_{\mu\nu,i}\Phi^i + \frac{\lambda^2}{2} B_{\mu\nu,ij}\Phi^i\Phi^j + \cdots \qquad (39.4)$$

$$C^{(n)} = \lambda C^{(n)}_{,i}\Phi^i + \frac{\lambda^2}{2} C^{(n)}_{,ij}\Phi^i\Phi^j + \cdots \qquad (40.4)$$

بخشی از اثر میدان dilaton را می‌توان با بازتعریف میدانهای دیگر حذف کرد. ما در این پایان‌نامه ساده‌سازی‌ای انجام می‌دهیم و کلاً این میدان را صفر می‌گذاریم. در حالت ایستا با جایگذاری بسطهای (36.4) / (39.4) و (40.4) در کنش تعدادی $D$-ذره (1.4) و بسط کنش بر حسب $\lambda$ و نگه‌داری جمله‌ها تنها تا مرتبه‌ی $\lambda^2$ به دست می‌آوریم

$$S = -T_0 \int d\tau \, Str( \ \frac{\lambda^2}{2} M_{ij}[\Phi^i, \Phi^j] + i\lambda^2 N_{ijl}[\Phi^i, \Phi^j]\Phi^l$$

$$+ \lambda^2 P_{ijkl}[\Phi^l, \Phi^k][\Phi^j, \Phi^i] + O(\lambda^3)) \qquad (41.4)$$

که کمیتهای $M_{ij}$ / $N_{ijk}$ و $P_{ijkl}$ بر حسب میدانهای زمینه به صورت زیر می‌باشند

$$M_{ij} = -\frac{1}{4} G_{00,ij} - \frac{1}{2} C^{(1)}_{0,ij} \qquad (42.4)$$

$$N_{ikl} = \frac{1}{2} B_{[ki,l]} - \frac{1}{2} C^{(3)}_{0[ki,l]} - \frac{1}{2} C^{(1)}_0 B_{[ki,l]} \qquad (43.4)$$

$$P_{ijkl} = \frac{1}{4} \delta_{kj}\delta_{li} + \frac{1}{8} C^{(5)}_{ijkl0} \qquad (44.4)$$

۴۸

این میدانهای بالا تقارنهای زیر را دارند

$$M_{ij} = M_{ji} \tag{۴۵.۴}$$

$$N_{ijl} = N_{[ijk]} \tag{۴۶.۴}$$

$$P_{ijkl} = -P_{jhkl} \tag{۴۷.۴}$$

$$P_{ijkl} = -P_{ijlk} \tag{۴۸.۴}$$

$$P_{[ijk]l} = ۰ \tag{۴۹.۴}$$

این کمیتها میدانهایی هستند که باعث مهگون شدن و درهم‌تنیده شدن $D$-ذره‌ها می‌شوند. پتانسیل سه‌فرمی که در بخش قبل مرورش کردیم باعث روشن شدن یکی از این کمیتها می‌شود و بستری برای درهم‌تنیده‌شدن $D$-ذره‌ها مهیا می‌کند. روشن کردن میدانهای غیریکنواخت متریک، پتانسیلهای فرمی‌ی بالاتر، میدان نو-شوارتز نیز می‌تواند $D$-ذره‌ها را قطبیده کنند و باعث درهم‌تنیدگی آنها شوند.

ما در ادامه‌ی این بخش دو جواب جدید ناجابه‌جایی را معرفی می‌کنیم.

## ۴.۲.۱ بیضی‌گون مهگون

در نبود $C^{(۵)}$ معادله‌ی حرکت کنش (۴۱.۴) عبارت می‌شود از

$$۲\lambda M_{in}\Phi^i + ۳i\lambda N_{nkl}[\Phi^k, \Phi^l] + \lambda[\Phi^i, [\Phi^i, \Phi^n]] = ۰ \tag{۵۰.۴}$$

که در نبود $M_{in}$ و $N_{nkl} = \tilde{N}\epsilon_{nkl}$ دقیقا رابطه‌ی (۱۵.۴) می‌باشد. در رابطه‌ی (۱۵.۴) جمله‌ی $\epsilon_{nkl}[\Phi^k, \Phi^l]$ تنها ناشی از پتانسیل سه‌فرم $C^{(۳)}$ بود. اما در رابطه‌ی (۵۰.۴) جمله‌ی $\epsilon_{nkl}[\Phi^k, \Phi^l]$ ناشی از میدانهای $B$ و $C^{(۳)}$ می‌باشد. در

۴۹

واقع داریم

$$\tilde{N} = \frac{1}{2}(C^{(3)}_{0123} - H_{123}) \tag{۵۱.۴}$$

که $H_{123}$ قدرت میدان $B$ است. پس در بودِ میدان $D$ $B$-ذرها به صورت یک دوقطبی‌ی الکتریکی می‌توانند مانند این میدان را احساس کند. این موضوع در [۱۰] هم گزارش شده است.

در حالتی که جمله‌ی $N_{ijk}$ صفر باشد و تنها جمله‌ی $M_{ij}$ وجود داشته باشد رابطه‌ی (۵۰.۴) تبدیل می‌شود به

$$2M_{in}\Phi^n + [\Phi^j, [\Phi^j, \Phi^i]] = 0 \tag{۵۲.۴}$$

با یک دوران بر روی $\Phi$ ها می‌توان $M_{ij}$ را قطری کرد و رابطه‌ی (۵۲.۴) را به رابطه‌ی زیر ساده کرد

$$2M_i\Phi^i + [\Phi^j, [\Phi^j, \Phi^i]] = 0 \tag{۵۳.۴}$$

که $M_i$ در آیه‌های ماتریس قطری‌ی $M_{ij}$ می‌باشند. در ادامه فرض می‌کنیم تنها متغیرهای ناجابه‌جایی عبارت باشند از $\Phi^1$، $\Phi^2$ و $\Phi^3$. به نام یک حدس به دنبال جواب‌هایی می‌گردیم که در جبر $SU(2)$ به صورت زیر صدق کنند:

$$\Phi^i = \frac{a^i}{2}\alpha^i \tag{۵۴.۴}$$

$$[\Phi^i, \Phi^j] = i\frac{a^i a^j}{a^k}\epsilon_{ijk}\Phi^k \tag{۵۵.۴}$$

$$tr(\Phi^i\Phi^j) = \delta^{ij}tr(\Phi^i\Phi^i) \tag{۵۶.۴}$$

دقت شود که در سمت راست معادله‌های (۵۴.۴) و (۵۵.۴) بر روی $i$ و $j$ جمع بسته نشده است. با جای‌گذاری‌ی (۵۵.۴)

۵۰

در (۵۳.۴) و استفاده (۵۶.۴) رابطه‌ی (۵۳.۴) تبدیل می‌شود به

$$2M_1 + a_2^2 + a_3^2 = 0 \qquad (57.4)$$

$$2M_2 + a_1^2 + a_3^2 = 0 \qquad (58.4)$$

$$2M_3 + a_1^2 + a_2^2 = 0 \qquad (59.4)$$

که به حل‌هایی برای $a$ها می‌انجامند

$$a_1^2 = M_1 - M_2 - M_3 \qquad (60.4\text{الف})$$

$$a_2^2 = M_2 - M_1 - M_3 \qquad (60.4\text{ب})$$

$$a_3^2 = M_3 - M_1 - M_2 \qquad (60.4\text{ج})$$

چون $a$ها حقیقی هستند، جواب‌های بالا شرطی بر روی $M$ها می‌گذارد. $M$هایی وجود دارند که این شرط را برآورده کنند. در فضای سه‌بعدی $M$ها، این شرط در یک هرم برآورده می‌شود. وجه‌های این هرم مربوط به وقتی است که یکی از $a$ها صفر باشند. برای $M$های درون یا روی این هرم، جواب ناجابه‌جایی به صورت زیر است

$$\Phi^1 = \frac{\alpha^1}{2}\sqrt{M_1 - M_2 - M_3} \qquad (61.4)$$

$$\Phi^2 = \frac{\alpha^2}{2}\sqrt{M_2 - M_1 - M_3} \qquad (62.4)$$

$$\Phi^3 = \frac{\alpha^3}{2}\sqrt{M_3 - M_2 - M_1} \qquad (63.4)$$

که در آن $\alpha^i$ ها نمایش $N \times N$ مولدهای هرمیتی‌ی جبر $SU(2)$ می‌باشند. با استفاده از (۲۰.۴) می‌توان دید که

$$\frac{(\Phi^1)^2}{a_1} + \frac{(\Phi^2)^2}{a_2} + \frac{(\Phi^3)^2}{a_3} = 4N(N^2 - 1)I_N \qquad (64.4)$$

اگر $\Phi$ ها عدد می‌بودند این رابطه یک بیضی‌گون را نمایش می‌داد. اکنون که ماتریس هستند رابطه‌ی (۶۴.۴) یک بیضی‌گون ناجابه‌جایی یا بیضی‌گون مه‌گون را نمایش می‌دهد. انرژی این بیضی‌گون مه‌گون عبارت می‌شود از

$$V(\Phi) = -T_\circ \frac{\lambda^۲}{۴}(a_۱^۲ a_۲^۲ + a_۲^۲ a_۳^۲ + a_۱^۲ a_۳^۲) tr((\alpha')^۲) \qquad (۶۵.۴)$$

چون این انرژی کمتر از صفر است $D$-ذره‌ها علاقه‌مند به واهلش به حالت درهم‌تنیده و مه‌گون هستند. در این مورد نیز نمایش کاهش ناپذیر $SU(N)$ پایین‌ترین انرژی را به ما می‌دهد.

میدان $M_{ij}$ ناشی از میدان متریک و پتانسیل تک فرم است. قسمت ناشی از متریک را می‌توان بر حسب تانسور ریمان نوشت:

$$g_{\circ\circ,ij} = ۲ R^\circ{}_{i\circ j} \qquad (۶۶.۴)$$

پس فضا-زمان خمیده باعث قطبش $D$-ذره‌ها می‌شود. میدان الکتریکی و مغناطیسی ناشی از پتانسیل تک فرم نیز در $M_{ij}$ وجود دارد. پس این میدان‌ها هم باعث قطبش $D$-ذره‌ها و در هم‌تنیدگی آنها می‌شوند.

## ۴.۲.۲ هذلولوی مه‌گون

در بخش پیش نشان دادیم که در حالتی که تنها $M_{ij}$ وجود داشته باشد و مقادیر ویژه‌ای این ماتریس در رابطه‌های زیر صدق کنند

$$M_۱ - M_۲ - M_۳ \leq ۰ \qquad (۶۷.۴)$$

$$M_۲ - M_۱ - M_۳ \leq ۰ \qquad (۶۸.۴)$$

$$M_۳ - M_۲ - M_۱ \leq ۰ \qquad (۶۹.۴)$$

آنگاه یک جواب نابدیهی با جبر $SU(2)$ و ماتریس‌هایی با بعد محدود برای $\Phi^i$ وجود دارد. اکنون می‌خواهیم جوابی برای حالت خاص $M_1 = 0$ و $M_2 = M_3 = M > 0$ بیابیم. این حالت آشکارا در شروط بالا صدق نمی‌کند. ما باید جوابی جدید پیدا کنیم. معادله‌های حرکت با این $M$ ها عبارت‌اند از

$$0 = [\Phi^2,[\Phi^1,\Phi^2]] + [\Phi^3,[\Phi^1,\Phi^3]] \tag{4.70}$$

$$2M\Phi^2 = [\Phi^1,[\Phi^2,\Phi^1]] + [\Phi^3,[\Phi^2,\Phi^3]] \tag{4.71}$$

$$2M\Phi^3 = [\Phi^1,[\Phi^3,\Phi^1]] + [\Phi^2,[\Phi^3,\Phi^2]] \tag{4.72}$$

به سادگی می‌توان دید که یک جواب معادله‌های بالا وقتی است که $\Phi$ ها جبر لی‌ی پایین را می‌سازند:

$$[\Phi^2,\Phi^3] = i\theta \tag{4.73}$$

$$[\Phi^1,\Phi^3] = i\sqrt{2M}\Phi^2 \tag{4.74}$$

$$[\Phi^1,\Phi^2] = i\sqrt{2M}\Phi^3 \tag{4.75}$$

که $\theta$ عددی ثابت است. عملگر کازیمیر جبر بالا عبارت است از

$$\Phi^1 - \frac{\sqrt{2M}}{2\theta}((\Phi^3)^2 - (\Phi^2)^2) = J \tag{4.76}$$

با یک انتقال در $\Phi^1$ عملگر $J$ را همیشه می‌توان صفر کرد و به دست آورد

$$\Phi^1 = \frac{\sqrt{2M}}{2\theta}((\Phi^3)^2 - (\Phi^2)^2) \tag{4.77}$$

۵۳

اگر $\Phi^1$، $\Phi^2$ و $\Phi^3$ عدد می‌بودند رابطه‌ی بالا یک هذلولوی را تعریف می‌کرد. اکنون که ماتریس هستند به رابطه‌ی بالا اصطلاحاً یک هذلولوی مه‌گون می‌گوییم.

انرژی پتانسیل هذلولوی مه‌گون عبارت می‌شود از

$$V \approx T_\circ M(tr((\Phi^2)^2 + (\Phi^3)^2)) \tag{۴.۷۸}$$

این انرژی مثبت است. پس این جواب فرین معادلات حرکت است اما لزوماً کمینه‌ی معادلات حرکت نیست. رابطه‌ی $[\Phi^2, \Phi^3] = i\theta$ با تبدیلات

$$\theta \rightarrow \hbar \tag{۴.۷۹}$$

$$\Phi^2 \rightarrow x \tag{۴.۸۰}$$

$$\Phi^3 \rightarrow p \tag{۴.۸۱}$$

تبدیل به رابطه‌ی آشنای $[x,p] = i\hbar$ می‌شود. با این تبدیل پتانسیل به صورت زیر خوانده می‌شود

$$V \approx MT_\circ(tr(x^2 + p^2)) \tag{۴.۸۲}$$

اگر رد را تبدیل به انتگرال کنیم. این انرژی یک نوسانگر هماهنگ ساده با طیف انرژی گسسته می‌شود. با تبدیل جمع بر روی پایه‌های انرژی، به دست می‌آوریم

$$V \approx 2\theta M T_\circ \sum_{n=\circ}^{\infty} n \tag{۴.۸۳}$$

اگر بی‌نهایت $D$-ذره داشته باشیم یک هذلولوی کامل خواهیم داشت. در صورتی که تعداد زیاد اما نابی‌نهایتی $D$-ذره داشته

۵۴

باشیم باید در عبارت بالا به جای $\infty$ تعداد محدود $D$-ذره‌ها را بگذاریم. با این کار به دست می‌آوریم

$$V \approx T_\circ M \theta N^2 \qquad (8.4.4)$$

پس میانگین برهم‌کنش $D$-ذره‌ها متناسب می‌شود با $T_\circ \theta M$. این انرژی برهم‌کنشی در درآیه‌های مختلف ماتریس‌ها پخش شده‌اند.

دقت کنید که ما در این بخش، به عنوان پخش پایانی این پایان‌نامه یک فرین مه‌گون نوین را برای $D$-ذره‌ها یافتیم. این فرین نو یک حالت برانگیخته‌ی $D$-ذره‌ها است. این برانگیختگی می‌تواند با ایجاد خمش مناسب در هندسه‌ی فضا-زمان یا روشن کردن میدان‌های مغناطیسی یا الکتریکی‌ی پتانسیل تک فرم روی دهد.

<p style="text-align: center;">کتاب‌نامه</p>